\documentclass[prd,aps,showpacs,epsf,floats,onecolumn]{revtex4}
\usepackage{amssymb}
\usepackage{amsfonts}
\usepackage{amsmath}
\usepackage{graphicx}

\begin{document}

\title{\textbf{Information Geometric Aspects of Probability Paths with
Minimum Entropy Production for Quantum State Evolution}}
\author{\textbf{Steven Gassner}$^{1}$, \textbf{Carlo Cafaro}$^{1}$, \textbf{%
Sean A. Ali}$^{2}$, \textbf{Paul M.\ Alsing}$^{3}$}
\affiliation{$^{1}$SUNY Polytechnic Institute, 12203 Albany, New York, USA}
\affiliation{$^{2}$Albany College of Pharmacy and Health Sciences, 12208 Albany, New
York, USA}
\affiliation{$^{3}$Air Force Research Laboratory, Information Directorate, 13441 Rome,
New York, USA}

\begin{abstract}
We present an information geometric analysis of both entropic speeds and
entropy production rates arising from geodesic evolution on manifolds
parametrized by pure quantum states. In particular, we employ pure states
that emerge as outputs of suitably chosen $\mathrm{su}\left( 2\text{; }%
\mathbb{C}
\right) $ time-dependent Hamiltonian operators that characterize analog
quantum search algorithms of specific types. The $\mathrm{su}\left( 2\text{; 
}%
\mathbb{C}
\right) $ Hamiltonian models under consideration are specified by external
time-dependent magnetic fields within which spin-$1/2$ test particles are
immersed. The positive definite Riemannian metrization of the parameter
manifold is furnished by the Fisher information function. The Fisher
information function is evaluated along parametrized squared probability
amplitudes obtained from the temporal evolution of these spin-$1/2$ test
particles. A minimum action approach is then utilized to induce the transfer
of the quantum system from its initial state to its final state on the
parameter manifold over a finite temporal interval. We demonstrate in an
explicit manner that the minimal (that is, optimum) path corresponds to the
shortest (that is, geodesic) path between the initial and final states.
Furthermore, we show that the minimal path serves also to minimize the total
entropy production occurring during the transfer of states. Finally, upon
evaluating the entropic speed as well as the total entropy production along
optimal transfer paths within several scenarios of physical interest in
analog quantum searching algorithms, we demonstrate in a transparent
quantitative manner a correspondence between a faster transfer and a higher
rate of entropy production. We therefore conclude that higher entropic speed
is associated with lower entropic efficiency within the context of quantum
state transfer.
\end{abstract}

\pacs{Information Theory (89.70.+c),
Probability
Theory
(02.50.Cw),
Quantum
Computation (03.67.Lx),
Quantum
Information
(03.67.Ac), Quantum
Mechanics
(03.65.-w), Riemannian
Geometry
(02.40.Ky),
Statistical Mechanics
(05.20.-y).}
\maketitle

\section{Introduction}

Understanding the intersection between thermodynamics and quantum mechanics
is becoming increasingly relevant, especially at the nanoscale level, since
applications such as quantum computing are moving from the theoretical arena
to real world applications. In particular, the necessity of designing
quantum algorithms that are not simply faster, but also optimized from a
thermodynamical standpoint, is becoming increasingly evident \cite%
{frey,nature,renner18, deffner17}. Because of this, it is insightful to
study algorithms such as analog quantum search algorithms using geometric
ideas, which are of great interest in both quantum mechanics and
thermodynamics alike.

The idea of fluctuations is central to the Riemannian geometrization of both
quantum mechanics and thermodynamics. In quantum mechanics, for instance, a
natural distance on the set of quantum states is represented by the angle in
Hilbert space. In particular, up to a constant factor, the angle is the only
Riemannian metric on the set of rays which is invariant under all possible
unitary time evolutions. More importantly perhaps, Wootters showed in Ref. 
\cite{wootters} that this natural distance is equivalent to the statistical
distance defined between different preparations of the same quantum system
and is determined completely by the size of statistical fluctuations
occurring in quantum measurements designed to distinguish one quantum state
from another. In quantum mechanics, statistical fluctuations are unavoidable
and are associated with a finite sample. In particular, two preparations are
indistinguishable in a given number of trials if the difference in the
actual probabilities is smaller than the size of a typical fluctuation. In
thermodynamics, including the theory of fluctuations in the axioms of
equilibrium thermodynamics \cite{landau}, a natural distance on the set of
equilibrium thermodynamic states is given in terms of a positive-definite
Riemannian metric proportional to the second order partial derivative of the
total entropy of the thermodynamic system and the reservoir with which it is
in contact \cite{ruppeiner79,ruppeiner95}. Roughly speaking, according to
the theory of fluctuations, when a thermodynamic system is in contact with a
reservoir, it reaches a steady-state condition in finite time by thermally
fluctuating among a continuous sequence of intermediate states. In
particular, the physical interpretation of distance between two
thermodynamic states can be described by stating that the less the
probability of fluctuation between the states, the further apart they are.
Therefore, in both quantum mechanics and thermodynamics, Riemannian
geometric structures emerge because of the presence of fluctuations.

Furthermore, although arising from a different physical origin, either
statistical fluctuations in quantum mechanics or thermal fluctuations in
thermodynamics are concepts that can be described quantitatively in terms of
probabilities. Indeed, both the thermodynamic \cite{salamon85,crooks07} and
the quantum \cite{caves94} metric tensors can be described in a more general
information-theoretic setting in terms of the Fisher information \cite%
{cover06}. This quantity is central to information geometry \cite{amari00},
the latter being the application of Riemannian geometry to probability
calculus \cite{felice18}. In particular, the Fisher information matrix is a
covariance matrix that is symmetric and positive semi-definite. Therefore,
it can be employed as a metric tensor to define a notion of distance between
points (for instance, probabilities, pure states, or density operators) in
the space of parameters.

In the framework of finite-time thermodynamic processes controlled by
external reservoirs \cite{salamon85,salamon83,salamon89,ries95}, Riemannian
geometric methods have been used to find the best path in the state space
along which one drives the thermodynamic system from a given equilibrium
state to another. Specifically, using the concepts of thermodynamic length
and thermodynamic divergence of a path, it was shown that the optimum
cooling paths (that is, paths of minimum entropy production) correspond to
geodesic motion with constant thermodynamic speed in the Riemann-metricized
manifold of thermodynamic states \cite{diosi96,diosi00}. In recent years,
the concept of thermodynamic length, adapted to the framework of nanoscale
processes \cite{crooks07,sivak12,crooks12b}, has been used in a number of
works to find optimal control parameter protocols in simple non-equilibrium
models. In this context, optimum paths are geodesics on the parameter
manifold and are non-equilibrium control protocols with lowest achievable
dissipation \cite{crooks15,crooks17}. Moreover, inspired by Jaynes'
information-theoretic reformulation of statistical mechanics \cite%
{jaynes57a,jaynes57b}, Frieden and collaborators discussed the role played
by the Fisher information in both equilibrium and non-equilibrium
thermodynamics in Refs. \cite{frieden98,frieden99,frieden02}.

Within the context of quantum search algorithms \cite{grover97,nielsenbook},
Riemannian geometric concepts have been employed to facilitate novel
modeling and analyses across a diverse range of differing approaches. In 
\cite{alvarez00} for instance, it was demonstrated that Grover's search
algorithm can be specified via a unitary, adiabatic process that leaves the
Fisher information function invariant. In \cite{wadati01} on the other hand,
the role of entanglement in quantum searching was investigated from the
perspective of the Fubini-Study metric. Further still, upon quantifying the
concept of quantum distinguishability among parametric density operators by
means of the Wigner-Yanase quantum information metric in \cite{cafaro2012A,
cafaro2012B}, it was demonstrated that the problem of quantum searching can
be reformulated in an information geometric setting, wherein Grover's
dynamics can be characterized by geodesic curves on the parametric manifold
of pure quantum state density operators. These density operators in turn,
are obtained from the continuous approximation of parametric quantum output
states of Grover's search algorithm. Finally, in Ref. \cite{cafaro2017},
information geometric techniques were used to confirm the superfluous nature
of the Walsh-Hadamard operation and, more significantly, to reproduce the
relation describing quadratic speed-up.

The thermodynamical perspectives on quantum computation and information
provided by \cite{bennett82} and \cite{parrondo15}\textbf{, }respectively,
together with that on quantum error correction provided by \cite%
{carlo13,carlo14b}, can prove deeply informative. When viewed in the context
of simple quantum circuit models, the performance of search schemes in the
framework of both classical and quantum algorithms is usually quantified in
terms of the so-called query complexity of the algorithm. By query
complexity we refer to the number of oracle queries implemented by the
algorithm. Increasingly realistic models of computation must be considered
in an effort to properly analyze quantum speed-ups. Among other aspects,
physically realistic analyses should take account of the thermodynamical
(resource) costs of implementing such algorithms on actual computer
hardware. Several initial results appear in \cite{beals13,perlner17}. By
making use of Bennett's Brownian model of reversible, low power computation 
\cite{bennett82} for instance, Perlner and Liu argued in Ref. \cite%
{perlner17} that exhaustive classical searching can be robustly competitive
with Grover's algorithm when the comparison between the two quantum
searching protocols is made in terms of actual thermodynamical (resource)
costs including memory size, energy consumption and computation time. The
comparative analysis presented in Ref. \cite{perlner17} focuses entirely
upon the comparison between Grover's quantum search algorithm and classical
search schemes. We remark that a comparison among alternative quantum search
protocols is absent in Ref. \cite{perlner17}.

An information geometric description of monotonic and oscillatory behavior
of (statistically) parametrized squared probability amplitudes arising from\
Fisher information functional forms that are selected \emph{ad hoc}, namely
constant, power-law and exponential decay, was presented in Ref. \cite%
{cafaropre18}. For each of these cases, the speed as well as thermodynamic
divergence of the associated physical processes\ was computed by utilizing a
useful Riemannian geometrization of relevant thermodynamical concepts.
Furthermore, we discussed in brief the possibility of leveraging our
proposed information geometric framework to characterize a compromise
between thermodynamic efficiency and speed in quantum searching protocols.
One primary limitation of the analysis presented in Ref. \cite{cafaropre18}
is the fact that the Fisher information functional forms were selected in an 
\emph{ad hoc} manner without prescribing their emergence from a well defined
underlying physical setting. Despite the mathematical generality however,
the predicted behaviors of the parametrized squared probability amplitudes
that arise from our information geometric modeling lacks any clear physical
interpretation. Motivated by the results of Byrnes and collaborators \cite%
{byrnes18}, we presented in Ref. \cite{alsing19} an in-depth analysis
describing the physical relationship between analytically soluble two-level
time-dependent quantum systems and quantum search Hamiltonians. In
particular, we calculated in an exact\ analytical manner the transition
probabilities from a source state to target state in several physical
scenarios featuring a spin-$1/2$ test particle immersed in an external
time-dependent magnetic field. To be more specific, we investigated both the
monotonic as well as the periodic oscillatory temporal behaviors of such
transition probabilities and moreover, investigated their analogy with
intrinsic features of both fixed-point and Grover-like quantum searching
protocols, respectively. Finally, we discussed from a physical perspective
the relationship that exists between the schedule of a quantum searching
algorithm in both adiabatic and non-adiabatic quantum evolutions, and the
control magnetic fields present in time-dependent driving Hamiltonians.

In this paper, driven by the absence of any thermodynamical analysis of
quantum searching protocols and building upon our previous research
presented in Refs. \cite%
{cafaropre18,alsing19,alsing19b,carlopre20,carloquantum20,steven20}, we
incorporate the Riemannian geometric concepts of speed and efficiency within
both thermodynamical and quantum mechanical settings, in an effort to
provide some theoretical perspective on the compromise between efficiency
and speed in terms of minimal entropy production channels arising from
quantum evolutions. In particular, we present an information geometric
characterization of entropy production rates and entropic speeds in geodesic
evolution occurring on statistical manifolds of parametrized quantum states,
which emerge as output states of su$\left( 2\text{; }%
\mathbb{C}
\right) $ Hamiltonian models that approximate various types of
continuous-time quantum search protocols. Before describing how the present
article is organized, we emphasize that our work being presented here is a
more in depth unabridged version of Ref. \cite{carlopre20} and is more
suitable for an audience with a broader interest in mathematical methods in
theoretical physics.

The layout of this article is as follows: in Section II, we present several
relevant and useful information geometric concepts.\ Among others, we direct
our attention to the notions of Fisher information, thermodynamic divergence
and thermodynamic length. In Section III, we clarify in precise terms how
the pure quantum states that we consider arise as outputs of suitably chosen 
$\mathrm{su}\left( 2\text{; }%
\mathbb{C}
\right) $ time-dependent Hamiltonian operators used to describe differing
types of analog quantum searching algorithms. In Section IV, we introduce
the positive definite (i.e. Riemannian) metrization of the parameter
manifold, where the latter is specified by the Fisher information functional
evaluated along parametrized squared probability amplitudes. Our concluding
remarks are presented in Section V.

\section{Preliminaries in Information Geometry}

In this section we introduce some preliminaries in the study of information
geometry, particularly the concepts of Fisher information, thermodynamic
length, and thermodynamic divergence.

\subsection{Fisher information}

\subsubsection{Fisher information in information theory}

Assume $X$ to be a discrete random variable with alphabet $\mathcal{X}$, and
let $p_{X}(x$; $\theta )=p_{x}(\theta )$ be the probability mass function
for $X$ that depends on a single parameter $\theta $. The Fisher information 
$\mathcal{F}(\theta )$ quantifies how much information $X$ contains about
the parameter $\theta $. In other words, it provides a lower bound on the
error in estimating the value of $\theta $ from the data in $X$. Formally,
in the single-parameter case, it is defined as the variance of the score $V%
\overset{\text{def}}{=}\partial _{\theta }\{\log [p_{x}(\theta )]\}$, namely 
\begin{equation}
\mathcal{F}\left( \theta \right) \overset{\text{def}}{=}\mathrm{var}\left( 
\mathrm{V}\right) =E_{\theta }\left[ \left\{ \partial _{\theta }\log \left[
p_{x}\left( \theta \right) \right] \right\} ^{2}\right] \text{,}  \label{fif}
\end{equation}%
where $E_{\theta }[\cdot ]$ denotes the expected value with respect to $%
p_{x}(\theta )$ and\textbf{\ }$\partial _{\theta }\overset{\text{def}}{=}%
\partial /\partial \theta $. As a side remark, we point out that the mean
value of the score is zero. Then, given any unbiased estimator $T(X)$ of the
parameter $\theta $, the Cramer-Rao inequality states that the mean-squared
error of $T(\theta )$ is lower bounded by the reciprocal of the Fisher
information \cite{cover06}, 
\begin{equation}
\mathrm{var}\left( T\right) \geq \frac{1}{\mathcal{F}\left( \theta \right) }%
\text{.}
\end{equation}

In the case where $p_{x}(\theta )$ depends on multiple parameters $\theta 
\overset{\text{def}}{=}\left( \theta ^{1}\text{,..., }\theta ^{M}\right) $,
the Fisher information is a matrix with elements $\mathcal{F}_{\alpha \beta
}(\theta )$, defined by 
\begin{equation}
\mathcal{F}_{\alpha \beta }\left( \theta \right) \overset{\text{def}}{=}%
E_{\theta }\left[ \left( \mathrm{V}_{\alpha }-\left\langle \mathrm{V}%
_{\alpha }\right\rangle \right) \left( \mathrm{V}_{\beta }-\left\langle 
\mathrm{V}_{\beta }\right\rangle \right) \right] \text{,}  \label{fifa}
\end{equation}%
where $\left\langle \mathrm{V}_{\alpha }\right\rangle \overset{\text{def}}{=}%
E_{\theta }\left[ \mathrm{V}_{\alpha }\right] $ and $\mathrm{V}_{\alpha }%
\overset{\text{def}}{=}\partial _{\alpha }\left\{ \log \left[ p_{x}\left(
\theta \right) \right] \right\} $ with $\partial _{\alpha }\overset{\text{def%
}}{=}\partial /\partial \theta ^{\alpha }$. Alternatively, Eq. (\ref{fifa})
can be recast as, 
\begin{equation}
\mathcal{F}_{\alpha \beta }\left( \theta \right) \overset{\text{def}}{=}%
\sum_{x\in \mathcal{X}}p_{x}\left( \theta \right) \frac{\partial \log \left[
p_{x}\left( \theta \right) \right] }{\partial \theta ^{\alpha }}\frac{%
\partial \log \left[ p_{x}\left( \theta \right) \right] }{\partial \theta
^{\beta }}\text{.}  \label{JIJ2}
\end{equation}%
with \textquotedblleft $\log $\textquotedblright\ denoting the natural
logarithm. Finally, we can extend this definition to the case of a
continuous random variable $X$ with alphabet $\mathcal{X}$ and probability
density function $\rho _{X}(x$; $\theta )$. Then the quantity $\mathcal{F}%
_{\alpha \beta }(\theta )$ in Eq. (\ref{JIJ2}) becomes \cite{cover06} 
\begin{equation}
\mathcal{F}_{\alpha \beta }\left( \theta \right) \overset{\text{def}}{=}%
\int_{\mathcal{X}}\rho _{X}\left( x\text{; }\theta \right) \frac{\partial
\log \left[ \rho _{X}\left( x\text{; }\theta \right) \right] }{\partial
\theta ^{\alpha }}\frac{\partial \log \left[ \rho _{X}\left( x\text{; }%
\theta \right) \right] }{\partial \theta ^{\beta }}dx\text{.}  \label{JIJ}
\end{equation}

The Fisher information is a central idea in not only the fields of
information theory \cite{cover06} and information geometry \cite{amari00},
but also in the geometric descriptions of both quantum mechanics \cite%
{caves94,brau996} and statistical mechanics \cite{crooks12}.

\subsubsection{Fisher information in quantum mechanics}

From a quantum mechanical viewpoint, consider an $N\overset{\text{def}}{=}%
2^{n}$-dimensional complex Hilbert space $\mathcal{H}_{2}^{n}$. Furthermore,
consider two neighboring normalized pure states $\left\vert \psi\left(
\theta\right) \right\rangle $ and $\left\vert \psi^{\prime }\left(
\theta\right) \right\rangle $ expanded with respect to an orthonormal basis $%
\left\{ \left\vert k\right\rangle \right\} $ with $1\leq k\leq N$,%
\begin{equation}
\left\vert \psi\right\rangle \overset{\text{def}}{=}\sum_{k=1}^{N}\sqrt{p_{k}%
}e^{i\varphi_{k}}\left\vert k\right\rangle \text{ and }\left\vert \psi
^{\prime}\right\rangle \overset{\text{def}}{=}\sum_{k=1}^{N}\sqrt{%
p_{k}+dp_{k}}e^{i\left( \varphi_{k}+d\varphi_{k}\right) }\left\vert
k\right\rangle \text{,}  \label{explicit}
\end{equation}
where $p_{k}=p_{k}\left( \theta\right) $ and $\varphi_{k}=\varphi_{k}\left(
\theta\right) $ with $\theta$ being a continuous real parameter. The quantum
distinguishability metric on this manifold of Hilbert space rays is given by
the Fubini-Study metric \cite{caves94},%
\begin{equation}
ds_{\text{FS}}^{2}\overset{\text{def}}{=}\left\{ \cos^{-1}\left[ \left\vert
\left\langle \psi^{\prime}|\psi\right\rangle \right\vert \right] \right\}
^{2}=g_{\alpha\beta}^{\left( \text{FS}\right) }\left( \theta\right)
d\theta^{\alpha}d\theta^{\beta}\text{,}  \label{fubini}
\end{equation}
where the Fubini-Study metric tensor components $g_{\alpha\beta}^{\left( 
\text{FS}\right) }$ are related to the Fisher-Rao metric tensor components $%
\mathcal{F}_{\alpha\beta}\left( \theta\right) $ by the relation%
\begin{equation}
g_{\alpha\beta}^{\left( \text{FS}\right) }\left( \theta\right) =\frac {1}{4}%
\left[ \mathcal{F}_{\alpha\beta}\left( \theta\right) +4\sigma
_{\alpha\beta}^{2}\left( \theta\right) \right] \text{.}  \label{fubini2}
\end{equation}
The quantities $\mathcal{F}_{\alpha\beta}\left( \theta\right) $ and $%
\sigma_{\alpha\beta}^{2}\left( \theta\right) $ in Eq. (\ref{fubini2}) are
formally defined as,%
\begin{equation}
\mathcal{F}_{\alpha\beta}\left( \theta\right) \overset{\text{def}}{=}%
\sum_{k=1}^{N}\frac{1}{p_{k}\left( \theta\right) }\frac{\partial p_{k}\left(
\theta\right) }{\partial\theta^{\alpha}}\frac{\partial p_{k}\left(
\theta\right) }{\partial\theta^{\beta}}\text{,}
\end{equation}
and,%
\begin{equation}
\sigma_{\alpha\beta}^{2}\left( \theta\right) \overset{\text{def}}{=}%
\sum_{k=1}^{N}\frac{\partial\varphi_{k}\left( \theta\right) }{\partial
\theta^{\alpha}}\frac{\partial\varphi_{k}\left( \theta\right) }{%
\partial\theta^{\beta}}p_{k}\left( \theta\right) -\left( \sum_{k=1}^{N}\frac{%
\partial\varphi_{k}\left( \theta\right) }{\partial\theta^{\alpha}}%
p_{k}\left( \theta\right) \right) \left( \sum_{k=1}^{N}\frac {%
\partial\varphi_{k}\left( \theta\right) }{\partial\theta^{\beta}}p_{k}\left(
\theta\right) \right) \text{,}  \label{variance}
\end{equation}
respectively. In what follows, we shall assume that the non-negative term $%
\sigma_{\alpha\beta}^{2}\left( \theta\right) $ in Eq. (\ref{variance})
related to the variance of the phase changes is equal to zero. This
assumption is valid by rephasing in a convenient manner the basis vectors
used to express the state $\left\vert \psi\left( \theta\right) \right\rangle 
$ (for details, we refer to Ref. \cite{caves94} and Ref. \cite{cafaropre18}%
). In particular, the rephasing procedure requires that $\mbox{Im}\left[
\left\langle \psi\left( \theta\right) |k\right\rangle \left\langle
k|d\psi_{\perp}\left( \theta\right) \right\rangle \right] =0$, for any $%
1\leq k\leq N$. The state $\left\vert d\psi_{\perp}\right\rangle \overset{%
\text{def}}{=}\left\vert d\psi\right\rangle -\left\langle
\psi|d\psi\right\rangle \left\vert d\psi\right\rangle $ is the projection of 
$\left\vert d\psi\right\rangle $ orthogonal to $\left\vert \psi\right\rangle 
$ where $\left\vert d\psi \right\rangle \overset{\text{def}}{=}\left\vert
\psi^{\prime}\right\rangle -\left\vert \psi\right\rangle $ while $\left\vert
\psi\right\rangle $ and $\left\vert \psi^{\prime}\right\rangle $ are defined
in Eq. (\ref{explicit}). In summary, for a convenient choice of the basis
vectors $\left\{ \left\vert k\right\rangle \right\} $ used for the
decomposition in Eq. (\ref{explicit}), $g_{\alpha\beta}^{\left( \text{FS}%
\right) }\left( \theta\right) $ becomes proportional to $\mathcal{F}%
_{\alpha\beta}\left( \theta\right) $ as evident from Eq. (\ref{fubini2}). As
mentioned earlier, we refer to Ref. \cite{caves94} for further details.

Interestingly, after some manipulations, the classical Fisher information in
Eq. (\ref{JIJ}) can be recast in terms of probability amplitudes $\sqrt {%
\rho_{X}\left( x\text{; }\theta\right) }$,%
\begin{equation}
\mathcal{F}_{\alpha\beta}^{\left( \text{classical}\right) }\left(
\theta\right) =4\int\partial_{\alpha}\sqrt{\rho_{X}\left( x\text{; }%
\theta\right) }\partial_{\beta}\sqrt{\rho_{X}\left( x\text{; }\theta\right) }%
dx\text{,}  \label{QMclassical}
\end{equation}
where $\partial_{\alpha}\overset{\text{def}}{=}\partial/\partial\theta
^{\alpha}$. Although quantum extensions of the Fisher information are not
unique, a simple generalization of Eq. (\ref{QMclassical}) occurs upon
replacing the probability density function $\rho_{X}\left( x\text{; }%
\theta\right) $ by the density operator $\rho_{\theta}\left( x\right) $ and
the integral by the trace \cite{luo03,luo06},%
\begin{equation}
\mathcal{F}_{\alpha\beta}^{\left( \text{QM}\right) }\left( \theta\right) =4%
\mathrm{tr}\left[ \partial_{\alpha}\sqrt{\rho_{\theta}\left( x\right) }%
\partial_{\beta}\sqrt{\rho_{\theta}\left( x\right) }\right] \text{.}
\label{WY}
\end{equation}
The metric tensor in\ Eq. (\ref{WY}) is known as the Wigner-Yanase metric
and is simply four times the Fubini-Study metric tensor. For a quantum
mechanical evolution specified by the unitary evolution operator $\mathcal{U}%
\overset{\text{def}}{=}e^{-\frac{i}{\hslash}\mathcal{H}t}$ with $\mathcal{H}%
\left( t\right) \overset{\text{def}}{=}-i\hslash\left( \partial _{t}\mathcal{%
U}^{\dagger}\right) \mathcal{U}$, the quantum Fisher information is
proportional to the dispersion of the Hermitian operator $\mathcal{H}$ \cite%
{boixo07,taddei13},%
\begin{equation}
\mathcal{F}^{\left( \text{QM}\right) }\left( t\right) =\frac{4}{\hslash^{2}}%
\left\langle \Delta\mathcal{H}^{2}\left( t\right) \right\rangle \text{.}
\label{qfish}
\end{equation}
In Eq. (\ref{qfish}), $\Delta\mathcal{H}\overset{\text{def}}{=}\mathcal{H}%
-\left\langle \mathcal{H}\right\rangle $ with $\left\langle \mathcal{H}%
\right\rangle $ denoting the expectation value of the operator $\mathcal{H}$
with respect to the initial state of the quantum system. While the
Fubini-Study metric is defined on the space of pure states, the
non-uniqueness of the quantum Fisher metric becomes especially meaningful
when considering quantum mixed states \cite{pires16}. In the framework of
mixed density operators, it is particularly convenient to introduce the
Bures line element $ds_{\text{Bures}}^{2}$ \cite{bures69},%
\begin{equation}
ds_{\text{Bures}}^{2}\overset{\text{def}}{=}2\left[ 1-\mathrm{F}\left( \rho%
\text{, }\rho+d\rho\right) \right] \text{.}  \label{bures}
\end{equation}
In Eq. (\ref{bures}), the quantity $\mathrm{F}\left( \rho\text{, }%
\sigma\right) \overset{\text{def}}{=}\mathrm{tr}\sqrt{\rho^{1/2}\sigma
\rho^{1/2}}$ denotes the so-called Uhlmann fidelity \cite{uhlmann76}. In
particular, we point out that $ds_{\text{Bures}}^{2}$ reduces to $ds_{\text{%
FS}}^{2}$ for pure states.

Interestingly, the Ising model is an ideal testing ground for the
convergence of both statistical inference and information geometric methods
applied to thermodynamical reasoning \cite{ali16}. Indeed, within the
quantum setting, a very interesting link between quantum information
geometry and thermodynamics arises in the study of the Bures metric on
thermal state manifolds corresponding to the quantum Ising model \cite%
{zanardi07},%
\begin{equation}
ds_{\text{Bures}}^{2}\overset{\text{def}}{=}\frac{1}{4}\left\langle \Delta 
\mathcal{H}^{2}\left( t\right) \right\rangle d\beta ^{2}=\frac{1}{4}%
T^{2}C_{v}d\beta ^{2}\text{.}  \label{zanardi}
\end{equation}%
In Eq. (\ref{zanardi}), $T$ is the temperature, $C_{v}$ is the specific
heat, and $\left\langle \cdot \right\rangle $ denotes the expectation value
with respect to thermal Gibbs states. For further details, we refer to Ref. 
\cite{zanardi07}.

\subsubsection{Fisher information in statistical mechanics}

After some straightforward manipulations of Eq. (\ref{JIJ}), it can be shown
that the Fisher metric tensor in Eq. (\ref{JIJ}) can be recast as \cite%
{caticha12},%
\begin{equation}
\mathcal{F}_{\alpha \beta }\left( \theta \right) \overset{\text{def}}{=}%
-\left( \frac{\partial ^{2}\mathcal{S}\left( \theta \text{, }\theta ^{\prime
}\right) }{\partial \theta ^{\prime \alpha }\partial \theta ^{\prime \beta }}%
\right) _{\theta ^{\prime }=\theta }=-E_{\theta }\left[ \partial _{\alpha
\beta }^{2}\log \rho _{X}\left( x\text{; }\theta \right) \right] \text{,}
\label{IT}
\end{equation}%
where $\mathcal{S}\left( \theta \text{, }\theta ^{\prime }\right) $ is the
so-called logarithmic relative entropy,%
\begin{equation}
\mathcal{S}\left( \theta \text{, }\theta ^{\prime }\right) \overset{\text{def%
}}{=}-\int_{\mathcal{X}}\rho _{X}\left( x\text{; }\theta \right) \log \left[ 
\frac{\rho _{X}\left( x\text{; }\theta \right) }{\rho _{X}\left( x\text{; }%
\theta ^{\prime }\right) }\right] dx\text{.}
\end{equation}%
For the sake of completeness, we remark that by defining the logarithmic
relative entropy by means of the natural logarithmic function, its numerical
value is expressed in units of \textquotedblleft \textrm{nats}%
\textquotedblright\ where $1$ \textrm{nat }$\overset{\text{def}}{=}\left[
\log (2)\right] ^{-1}\mathrm{bits}$. Therefore, the temporal entropy rate
would be measured in \textrm{nats}/sec. Furthermore, from a statistical
mechanical standpoint, the thermodynamic metric tensor $g_{\alpha \beta
}^{\left( \text{thermo}\right) }\left( \theta \right) $ is defined as \cite%
{crooks07}\textbf{,}%
\begin{equation}
g_{\alpha \beta }^{\left( \text{thermo}\right) }\left( \theta \right) 
\overset{\text{def}}{=}\frac{\partial ^{2}\psi }{\partial \theta ^{\alpha
}\partial \theta ^{\beta }}=E_{\theta }\left[ \left( X_{\alpha
}-\left\langle X_{\alpha }\right\rangle \right) \left( X_{\beta
}-\left\langle X_{\beta }\right\rangle \right) \right] \text{,}
\label{Thermo}
\end{equation}%
where $\psi \overset{\text{def}}{=}\log \left( \mathcal{Z}\right) $ is the
Massieu thermodynamic potential (that is, the free entropy) with $\mathcal{Z}
$ being the partition function of the particular statistical mechanical
system being considered. To understand the identity between $\mathcal{F}%
_{\alpha \beta }\left( \theta \right) $ in\ Eq. (\ref{IT}) and $g_{\alpha
\beta }^{\left( \text{thermo}\right) }\left( \theta \right) $ in\ Eq. (\ref%
{Thermo}), consider a probability distribution given by the Gibbs
equilibrium ensemble%
\begin{equation}
\rho _{X}\left( x\text{; }\theta \right) \overset{\text{def}}{=}\exp \left[
c\left( x\right) +\sum_{i=1}^{M}\theta ^{i}X_{i}\left( x\right) -\psi \left(
\theta \right) \right] \text{,}  \label{Gibbs}
\end{equation}%
with $c\left( x\right) $ and $X_{i}\left( x\right) $ being smooth functions
on $\mathcal{X}$. Substituting Eq. (\ref{Gibbs}) into Eq. (\ref{Thermo}), we
obtain%
\begin{equation}
g_{\alpha \beta }^{\left( \text{thermo}\right) }\left( \theta \right)
=-E_{\theta }\left[ \partial _{\alpha \beta }^{2}\log \rho _{X}\left( x\text{%
; }\theta \right) \right] \text{.}  \label{genny}
\end{equation}%
Therefore, from Eqs. (\ref{genny}) and (\ref{IT}), we conclude that the
thermodynamic metric tensor is identical to the Fisher metric tensor \cite%
{crooks07}. Indeed, the justification of this identity is rooted in the fact
that the statistical definition of entropy is more general than the original
thermodynamic definition. For an overview of the physical meaning of the
Fisher information in information theory, quantum mechanics, and
thermodynamics, we refer to Table I.

\begin{table}[t]
\centering
\begin{tabular}{c|c|c|c|c}
\hline\hline
Theoretical Arena & Parameter of Interest & Fluctuating Observable & Fisher
Information & Length \\ \hline
information theory & elapsed time & score function & variance of the score & 
entropic \\ 
thermodynamics & temperature & energy & size of energy fluctuations & 
thermodynamic \\ 
quantum theory & magnetic field intensity & Hermitian operator & dispersion
of the operator & statistical \\ \hline
\end{tabular}%
\caption{Illustrative representation of common parameters of interest and
fluctuating observables together with the physical interpretation of Fisher
information and length through parameter space in information theory,
thermodynamics, and quantum theory.}
\end{table}

\subsection{Thermodynamic length and divergence}

We have seen that given a space of states, we can employ a suitable
Riemannian metric tensor to define the distance between states and thus the
length of paths between states. Points in this state space may represent
statistical distributions (as in Eq. (\ref{JIJ})), quantum state vectors (as
in Eq. (\ref{fubini2})), or thermodynamic states (as in Eq. (\ref{Thermo})).
A generalized concept that provides a convenient parametrization for these
types of spaces is the information-theoretic notion of entropy, which allows
one to use the Fisher metric tensor as a metric on the state space, such as
in Eq. (\ref{IT}). We have seen that this perspective not only allows one to
consider parameters more general than those typically used in
thermodynamics, but also makes transparent the connections between state
spaces in these different fields.

Riemannian metrics on thermodynamic equilibrium state spaces can be achieved
in a number of ways. Weinhold's approach \cite{weinhold75} used the second
derivatives of internal energy with respect to extensive variables, while
Ruppeiner's approach \cite{ruppeiner79} used the second derivatives of the
entropy as a function of extensive variables. These extensive variables
could be, for example, volume or mole number. These approaches were shown to
be conformally equivalent by Salamon and collaborators in Ref. \cite%
{salamon84}. Using the energy approach, Salamon and Berry defined in Ref. 
\cite{salamon83} the length of a path $\gamma _{\theta }$ in the space of
thermal states as follows, 
\begin{equation}
\mathcal{L}\left( \tau \right) \overset{\text{def}}{=}\int_{0}^{\tau }\sqrt{%
\frac{d\theta ^{\alpha }}{d\xi }g_{\alpha \beta }\left( \theta \right) \frac{%
d\theta ^{\beta }}{d\xi }}d\xi \text{,}  \label{length}
\end{equation}%
where the parameter $\theta $ depends on an affine parameter $\xi $ with $%
0\leq \xi \leq \tau $, and $g_{\alpha \beta }(\theta )$ is the thermodynamic
metric tensor introduced in Eq. (\ref{Thermo}). The quantity $\mathcal{L}%
(\tau )$, with physical dimensions of $(\text{energy})^{1/2}$, is called
the\ \textquotedblleft thermodynamic length\textquotedblright\ of the path $%
\gamma _{\theta }$. Similarly, one can define this length using the entropy
approach, in which case the \textquotedblleft entropy\textquotedblright\
lengths would differ from the \textquotedblleft energy\textquotedblright\
lengths by a factor of the square root of some average temperature
throughout the thermodynamic process under consideration \cite%
{salamon83,salamon84}. The length computed using the thermodynamic entropy
approach in the space of extensive variables can also be shown, under
suitable working conditions, to be equal to the length computed using the
entropy of a probability distribution in the space of parameters for the
distribution \cite{salamon85}. Along with the thermodynamic length of a
path, it is also useful to introduce the so-called thermodynamic divergence $%
\mathcal{I}(\tau )$ of $\gamma _{\theta }$, defined as \cite{crooks07}, 
\begin{equation}
\mathcal{I}\left( \tau \right) \overset{\text{def}}{=}\tau \int_{0}^{\tau }%
\frac{d\theta ^{\alpha }}{d\xi }g_{\alpha \beta }\left( \theta \right) \frac{%
d\theta ^{\beta }}{d\xi }d\xi \text{.}  \label{divergence}
\end{equation}%
To better understand the meaning of Eqs. (\ref{length}) and (\ref{divergence}%
), we note that by the Cauchy-Schwarz inequality applied to two smooth
functions $f_{1}(t)$ and $f_{2}(t)$, 
\begin{equation}
\left[ \int_{0}^{\tau }f_{1}\left( t\right) f_{2}\left( t\right) dt\right]
^{2}\leq \left\{ \int_{0}^{\tau }\left[ f_{1}\left( t\right) \right]
^{2}dt\right\} \left\{ \int_{0}^{\tau }\left[ f_{2}\left( t\right) \right]
^{2}dt\right\} \text{,}
\end{equation}%
it follows that $\mathcal{I}\geq \mathcal{L}^{2}$. That is, the
thermodynamic divergence is always greater than the square of the
thermodynamic length. The special case $\mathcal{I}=\mathcal{L}^{2}$ occurs
when the integrand in Eq. (\ref{divergence}) is constant along the path $%
\gamma _{\theta }$. The divergence $\mathcal{I}$ is a measure of the amount
of natural fluctuations along a path, while the length $\mathcal{L}$ is a
measure of the cumulative root-mean-square deviations along the path \cite%
{crooks07}. Given these facts, and given the connection between metric
tensors and statistical fluctuations such as in Eqs. (\ref{fifa}) and (\ref%
{Thermo}), we may view these quantities as control measures of dissipation
in finite-time thermodynamic processes, making them very useful in the
Riemannian geometric formulation of thermodynamics.

The Riemannian geometric formulation of thermodynamics concerns the problem
of reversibly transferring a thermodynamic system from some initial state to
some final state under the constraint of minimum total entropy production
(or minimum loss of availability). In particular, one considers a relaxation
process in which the system is driven by a sequence of $\tilde{n}\overset{%
\text{def}}{=}\tau /(\delta t)$ steps. Interestingly, when considering this
problem in terms of the thermodynamic divergence along a path $\gamma
_{\theta }$ defined in Eq. (\ref{divergence}), it can be shown that this
quantity equals the total entropy production (dissipation) of the path in
the continuum limit of a sequence of step equilibrations \cite{salamon85,
crooks07, diosi96}. That is, in the limit that $\delta t$ approaches zero
(and the number of steps $\tilde{n}$ approaches infinity), the process is
ideally reversible. This has found use in the study of control parameter
protocols that minimize dissipation in molecular machines, in which the
Riemannian manifold is induced by a friction tensor and the thermodynamic
divergence is proportional to the average excess work performed during the
protocol \cite{sivak12, crooks17}.

\subsection{Optimum paths: Speed and Efficiency}

The problem is therefore to find the optimum paths $\gamma _{\theta }$ that
minimize an action functional given by the length $\mathcal{L}$ in Eq. (\ref%
{length}). Let $\theta \overset{\text{def}}{=}\{\theta ^{\alpha }\}$, $1\leq
\alpha \leq M$ constitute the parameter space, where $M$ is the
dimensionality of the space, and let $\gamma _{\theta }$ be defined by $%
\theta (\xi )$ with $0\leq \xi \leq \tau $. Then the minimum length path
occurs under some optimal affine time parametrization $\xi $ (unique up to a
change of scale and origin). Quantitatively, by using variational calculus
and tensor algebra, requiring that $\delta \mathcal{L}=0$ and imposing the
requirement that $\delta \theta ^{\alpha }=0$ at the extrema of the path
leads to the familiar geodesic equations, 
\begin{equation}
\frac{d^{2}\theta ^{\alpha }}{d\xi ^{2}}+\Gamma _{\beta \gamma }^{\alpha }%
\frac{d\theta ^{\beta }}{d\xi }\frac{d\theta ^{\gamma }}{d\xi }=0\text{,}
\label{ge}
\end{equation}%
where the quantities $\Gamma _{\beta \gamma }^{\alpha }$ are the Christoffel
connection coefficients of the second kind, defined as \cite{felice90}, 
\begin{equation}
\Gamma _{\beta \gamma }^{\alpha }\overset{\text{def}}{=}\frac{1}{2}g^{\alpha
\delta }\left( \partial _{\beta }g_{\delta \gamma }+\partial _{\gamma
}g_{\beta \delta }-\partial _{\delta }g_{\beta \gamma }\right) \text{,}
\label{connection}
\end{equation}%
with $\partial _{\beta }\overset{\text{def}}{=}\partial /\partial \theta
^{\beta }$. By considering the Fisher information matrix elements $\mathcal{F%
}_{\alpha \beta }(\theta )$ in Eq. (\ref{fifa}) as the components of the
metric tensor $g_{\alpha \beta }(\theta )$, the geodesic equations in Eq. (%
\ref{ge}) become, 
\begin{equation}
\frac{d^{2}\theta ^{\alpha }}{d\xi ^{2}}+\frac{1}{2}\mathcal{F}^{\alpha
\beta }\left( \theta \right) \left[ \partial _{\nu }\mathcal{F}_{\beta \rho
}\left( \theta \right) +\partial _{\rho }\mathcal{F}_{\nu \beta }\left(
\theta \right) -\partial _{\beta }\mathcal{F}_{\nu \rho }\left( \theta
\right) \right] \frac{d\theta ^{\nu }}{d\xi }\frac{d\theta ^{\rho }}{d\xi }=0%
\text{.}  \label{general}
\end{equation}%
Interestingly, if we decide instead to minimize the thermodynamic divergence 
$\mathcal{I}(\tau )$ under the same working assumptions, the geodesic
equations become, 
\begin{equation}
\frac{d}{d\xi }\left[ \mathcal{F}_{\alpha \rho }\left( \theta \right) \frac{%
d\theta ^{\alpha }}{d\xi }\right] -\frac{1}{2}\frac{d\theta ^{\alpha }}{d\xi 
}\frac{\partial \mathcal{F}_{\alpha \beta }\left( \theta \right) }{\partial
\theta ^{\rho }}\frac{d\theta ^{\beta }}{d\xi }=0\text{,}  \label{general1}
\end{equation}%
which can readily be shown to yield identical geodesic paths to the
equations in Eq. (\ref{general}). That is to say, optimal paths minimizing
the thermodynamic length are also paths that minimize the thermodynamic
divergence. We also remark that Eq. (\ref{general1}) is the information
geometric analogue of Eq. ($36$) in Ref. \cite{diosi96} and Eq. ($6$) in
Ref. \cite{crooks17}. By exploiting the concept of geodesic paths, we can
define the \textquotedblleft thermodynamic\textquotedblright\ speed
(henceforth referred\textbf{\ }to as entropic speed $v_{E}$) as 
\begin{equation}
v_{\text{\textrm{E}}}\overset{\text{def}}{=}\sqrt{\frac{d\theta ^{\alpha }}{%
d\xi }g_{\alpha \beta }\left( \theta \right) \frac{d\theta ^{\beta }}{d\xi }}%
\text{,}  \label{vthermo}
\end{equation}%
which is constant along these optimum paths. Optimum paths also have a
constant rate of entropy production $r_{E}$, given by 
\begin{equation}
r_{\text{\textrm{E}}}\overset{\text{def}}{=}\frac{d\mathcal{I}}{d\tau }\text{%
,}  \label{EPR}
\end{equation}%
where $\mathcal{I}(\tau )$ is evaluated along an optimal path. This quantity
is identical to the squared invariant norm of the speed $v_{E}$ when
considering an optimal path. We use the term \textquotedblleft
entropic\textquotedblright\ speed since the statistical definition of
entropy is more general than the original thermodynamic definition, as
previously discussed. We will therefore refer to lengths and divergences as
\textquotedblleft entropic\textquotedblright\ quantities as well in future
discussions.

We emphasize that $v_{E}$ in Eq. (\ref{vthermo}) has analogues in both
thermodynamics and quantum mechanics. In thermodynamics, the thermodynamic
speed $v_{\text{thermo}}$ is defined as the natural-time derivative of the
thermodynamic length $\mathcal{L}$ computed from the entropy version of the
metric tensor, and is given by \cite{salamon88}, 
\begin{equation}
v_{\text{thermo}}\overset{\text{def}}{=}\frac{d\mathcal{L}}{d\xi }\text{,}
\end{equation}%
where the dimensionless natural time scale $\xi $ (sometimes referred to as
thermodynamic time) is defined in terms of the clock time $t$, the
temperature $T$ of the system, and the relaxation time $\epsilon (T)$ as
follows, 
\begin{equation}
d\xi \overset{\text{def}}{=}\frac{dt}{\epsilon \left( T\right) }\text{.}
\end{equation}%
In quantum mechanics on the other hand, the statistical speed $v_{\text{stat}%
}$ is defined as the rate of change of the absolute statistical distance $%
l(\theta )$ between two states (either pure states in Hilbert space or mixed
states in the space of density operators) along the path parametrized by $%
\theta $, given by \cite{wootters, caves94}, 
\begin{equation}
v_{\text{stat}}\overset{\text{def}}{=}\frac{dl\left( \theta \right) }{%
d\theta }\text{,}  \label{sspeed}
\end{equation}%
where $dl^{2}(\theta )\overset{\text{def}}{=}g_{\alpha \beta }^{(\text{FS}%
)}(\theta )d\theta ^{\alpha }d\theta ^{\beta }$ with $g_{\alpha \beta }^{(%
\text{FS})}(\theta )$ given by Eq. (\ref{fubini2}). The absolute statistical
distance $l(\theta )$ is interpreted as the maximum number of
distinguishable quantum states along the path parametrized by $\theta $
optimized over all possible generalized quantum measurements (that is, under
the most discriminating measurements). The\textbf{\ }Eq. (\ref{sspeed}) also
provides an intuitive physical interpretation for the quantum Fisher
information as the square root of the statistical speed \cite%
{taddei13,andresen94,pezze09}.

In the final paragraph of this section, we propose an information geometric
notion of entropic efficiency. We recall that the thermal efficiency $\eta _{%
\text{thermo}}$ of a heat engine in thermodynamics is defined as \cite%
{beretta05},%
\begin{equation}
\eta _{\text{thermo}}\overset{\text{def}}{=}1-\frac{Q_{\text{out}}}{Q_{\text{%
in}}}\text{.}  \label{tefficiency}
\end{equation}%
In Eq. (\ref{tefficiency}), $Q_{\text{out}}$ and $Q_{\text{in}}$ denote the
output and input thermal energies with $W_{\text{out}}\overset{\text{def}}{=}%
Q_{\text{in}}-Q_{\text{out}}$ being the actual work performed by the heat
engine. We also point out that heat resistance, friction, internal losses,
and heat leakage are some factors that can cause deviations from ideality in
real heat engines. For a minimum entropy production analysis in a heat
engine subject to thermal-resistance losses, we refer to Ref. \cite%
{salamon80A}. Furthermore, for recent important works on the existence of a
universal trade-off between efficiency and power in heat engines, we refer
to Refs. \cite{tasaki16,seifert18}. Finally, the efficiency of a quantum
evolution in the Riemannian approach to quantum mechanics is defined as \cite%
{anandan90},%
\begin{equation}
\eta _{\text{QM}}\overset{\text{def}}{=}1-\frac{\Delta s}{s}\text{,}
\label{qefficiency}
\end{equation}%
with $0\leq \eta _{\text{QM}}\leq 1$ and $\Delta s\overset{\text{def}}{=}%
s-s_{0}$. The quantity $s_{0}$ denotes the dimensionless distance along the
shortest geodesic $\gamma _{\text{shortest}}$ joining the fixed initial ($%
\left\vert \psi _{i}\right\rangle $) and final ($\left\vert \psi
_{f}\right\rangle $) points of the evolution that are distinct points on the
complex projective Hilbert space. The quantity $s$ instead, is the distance
along a given path $\gamma $ connecting $\left\vert \psi _{i}\right\rangle $
and $\left\vert \psi _{f}\right\rangle $ and is measured by the Fubini-Study
metric. We remark that $s$ does not depend on the specific Hamiltonian used
to transport the quantum state along the given path. It depends only on the
unparametrized curve $\gamma $ in the projective Hilbert space that is
determined by the quantum evolution. The quantum evolution is maximally
efficient when the evolution occurs with minimum time-energy uncertainty.
This scenario is characterized by $\eta _{\text{QM}}=1$ and is actualized
when $\gamma $ and $s$ approach $\gamma _{\text{shortest}}$ and $s_{0}$,
respectively.

In analogy to these definitions of efficiency in thermodynamics and quantum
mechanics, we find it useful to define entropic efficiency based on the
notion of minimum entropy production as opposed to minimum energy
dispersion. Given an initial and final point on an information manifold,
there is a set of distinct evolutions with entropy production rates $%
r_{E}^{(i)}\in \mathbb{R}_{+}\backslash \{0\}$, where the index
\textquotedblleft $i$\textquotedblright\ labels each evolution. We define
the entropic efficiency of a particular evolution between those two points
as, 
\begin{equation}
\eta _{E}\overset{\text{def}}{=}1-\frac{r_{E}}{r}\text{,}  \label{efficiency}
\end{equation}%
where $r\overset{\text{def}}{=}\left\{ \left\lceil r_{E}^{(i)}\right\rceil
\right\} $ is the maximum of the ceiling functions of the entropy production
rates of the different paths. By ceiling function, we mean the function
mapping $x\in \mathbb{R}$ to the least integer greater than or equal to $x$.
The quantity $r$ is a normalizing factor that can be interpreted as the
least integer upper bound of the entropy production rate of the hottest
among all cool paths available for the evolution between two points. It
ensures that the quantity $\eta _{E}$ has the key properties of being
adimensional and satisfying $0\leq \eta \leq 1$. The ideal value $\eta =1$
is achieved when the evolution corresponds to a path whose entropy
production remains ideally constant, and is therefore maximally cooled or
maximally reversible. We remark that in such a case, just as the quantum
mechanical efficiency $\eta _{\text{QM}}$ approaches $1$ as $\Delta
s\rightarrow 0$, so too does the quantity $\eta _{E}$ approach $1$ as $%
r_{E}\rightarrow 0$.

Building on the preliminary concepts introduced in this section, we will
describe in the next section the particular quantum mechanical evolutions
that generate the probability paths to which we will apply these notions of
speed and minimum entropy production from an information geometric viewpoint.

\section{\textrm{su}$(2$; $%
\mathbb{C}
)$ Hamiltonian models}

In this section, we show how the pure states we consider can result from
evolution under a suitably defined $\mathrm{su}(2\text{; }\mathbb{C})$
time-dependent Hamiltonian operator. These pure states are useful for
understanding various types of analog quantum search strategies using
physical reasoning \cite{alsing19,alsing19b}.

\subsection{Hamiltonian evolution}

We start with a time-dependent Hamiltonian operator $\mathcal{H}_{\mathrm{su}%
(2\text{; }\mathbb{C})}(t)$ defined in terms of three anti-Hermitian and
traceless generators of $\mathrm{su}(2\text{; }\mathbb{C})$, the Lie algebra
of the special unitary group $\mathrm{SU}(2\text{; }\mathbb{C})$ \cite%
{sakurai94}. These generators are $\{i\sigma _{x}\text{, }-i\sigma _{y}\text{%
, }i\sigma _{z}\}$, where $\vec{\sigma}\overset{\text{def}}{=}(\sigma _{x}%
\text{, }\sigma _{y}\text{, }\sigma _{z})$ is the familiar Pauli vector
operator \cite{carlopra10,carlopra14}. We write $\mathcal{H}_{\mathrm{su}(2%
\text{; }\mathbb{C})}(t)$ as a linear superposition of these generators in
terms of complex, time-dependent coefficients ($a(t)$, $b(t)$, $c(t)$ $\in 
\mathbb{C}$), 
\begin{equation}
\mathcal{H}_{\mathrm{su}\left( 2\text{; }%
\mathbb{C}
\right) }\left( t\right) \overset{\text{def}}{=}a\left( t\right) \left(
i\sigma _{x}\right) +b\left( t\right) \left( -i\sigma _{y}\right) +c\left(
t\right) \left( \text{ }i\sigma _{z}\right) \text{.}  \label{peter}
\end{equation}%
If we set $a(t)\overset{\text{def}}{=}-i\omega _{x}(t)$, $b(t)\overset{\text{%
def}}{=}i\omega _{y}(t)$, and $c(t)\overset{\text{def}}{=}-i\Omega (t)$, we
have, 
\begin{equation}
\mathcal{H}_{\mathrm{su}\left( 2\text{; }%
\mathbb{C}
\right) }\left( t\right) \overset{\text{def}}{=}\omega _{x}\left( t\right)
\sigma _{x}+\omega _{y}\left( t\right) \sigma _{y}+\Omega \left( t\right)
\sigma _{z}\text{.}  \label{peter2}
\end{equation}%
Defining $\omega (t)\overset{\text{def}}{=}\omega _{x}(t)-i\omega
_{y}(t)=\omega _{\mathcal{H}}(t)\,e^{i\phi _{\omega }(t)}$, we have recast
the Hamiltonian in terms of the so-called transverse field $\omega (t)$ and
longitudinal field $\Omega (t)$. The transverse field is a complex quantity
oriented in the $xy$-plane, while the longitudinal field is a real quantity
oriented along the $z$-axis. These definitions cause $\mathcal{H}_{\mathrm{su%
}(2\text{; }\mathbb{C})}(t)$ to bear resemblance to the Hamiltonian of a
spin-$1/2$ particle in a time-dependent external magnetic field. For
instance, considering an electron with magnetic moment $\vec{\mu}\overset{%
\text{def}}{=}\mu _{\text{Bohr}}\,\vec{\sigma}$ in an external magnetic
field $\vec{B}(t)$, we can recast Eq. (\ref{peter2}) as 
\begin{equation}
\mathcal{H}_{\mathrm{su}\left( 2\text{; }%
\mathbb{C}
\right) }\left( t\right) \overset{\text{def}}{=}-\vec{\mu}\cdot \vec{B}%
\left( t\right) \text{.}  \label{peter3}
\end{equation}%
We denote by $\mu _{\text{Bohr}}\overset{\text{def}}{=}e\hslash /(2mc)$ the
Bohr magneton which is defined in terms of the electron mass $m$, the
elementary charge $e$, the speed of light $c$, and the reduced Planck
constant $\hslash $. We can then find the fields $\omega (t)$ and $\Omega
(t) $ in terms of the components of the external magnetic field $\vec{B}(t)$
using the following decomposition, 
\begin{equation}
\vec{B}\left( t\right) \overset{\text{def}}{=}\vec{B}_{\perp }\left(
t\right) +\vec{B}_{\parallel }\left( t\right) \text{,}  \label{peter4}
\end{equation}%
where $\vec{B}_{\perp }\left( t\right) \overset{\text{def}}{=}B_{x}\left(
t\right) \hat{x}+B_{y}\left( t\right) \hat{y}$ and $\vec{B}_{\parallel
}\left( t\right) \overset{\text{def}}{=}B_{z}\left( t\right) \hat{z}$. It is
then clear that 
\begin{equation}
B_{x}\left( t\right) =-\frac{2mc}{e\hslash }\omega _{x}\left( t\right) \text{%
, }B_{y}\left( t\right) =-\frac{2mc}{e\hslash }\omega _{y}\left( t\right) 
\text{, and }B_{z}\left( t\right) =-\frac{2mc}{e\hslash }\Omega \left(
t\right) \text{,}
\end{equation}%
or, in terms of the field magnitudes $\omega _{\mathcal{H}}(t)$ and $\Omega
_{\mathcal{H}}(t)$, 
\begin{equation}
B_{\perp }\left( t\right) =\frac{2mc}{\left\vert e\right\vert \hslash }%
\omega _{\mathcal{H}}\left( t\right) \text{, and }B_{\parallel }\left(
t\right) =\frac{2mc}{\left\vert e\right\vert \hslash }\Omega _{\mathcal{H}%
}\left( t\right) \text{.}  \label{xxx}
\end{equation}

\subsection{Transition Probabilities}

We now turn to the highly nontrivial task of obtaining exact analytical
expressions for the transition probability from an initial source state $%
|s\rangle $ to a final target state $|w\rangle $ under quantum mechanical
evolution specified by the Hamiltonian in Eq. (\ref{peter3}). Consider the
unitary evolution operator $\mathcal{U}(t)$, satisfying $i\hslash \,\dot{%
\mathcal{U}}(t)=\mathcal{H}_{\mathrm{su}(2\text{; }\mathbb{C})}(t)\mathcal{U}%
(t)$ with $\dot{\mathcal{U}}\overset{\text{def}}{=}\partial _{t}\mathcal{U}$%
. For convenience, we choose a basis $\{|w\rangle \text{, }|w_{\perp
}\rangle \}$ such that $\sigma _{z}|w\rangle =+|w\rangle $ and $\sigma
_{z}|w_{\perp }\rangle =-|w_{\perp }\rangle $, and write the matrix
representations of $\mathcal{H}_{\mathrm{su}(2\text{; }\mathbb{C})}(t)$ and $%
\mathcal{U}(t)$ in this basis as follows, 
\begin{equation}
\left[ \mathcal{H}_{\mathrm{su}\left( 2\text{; }%
\mathbb{C}
\right) }\right] \overset{\text{def}}{=}\left( 
\begin{array}{cc}
\Omega \left( t\right) & \omega \left( t\right) \\ 
\omega ^{\ast }\left( t\right) & -\Omega \left( t\right)%
\end{array}%
\right) \text{, and }\left[ \mathcal{U}\left( t\right) \right] \overset{%
\text{def}}{=}\left( 
\begin{array}{cc}
\alpha \left( t\right) & \beta \left( t\right) \\ 
-\beta ^{\ast }\left( t\right) & \alpha ^{\ast }\left( t\right)%
\end{array}%
\right) \text{.}  \label{evolutiono}
\end{equation}%
Note that $\mathcal{U}(t)$ is expressed in terms of two probability
amplitudes $\alpha (t)$ and $\beta (t)$, which by the unitarity of $\mathcal{%
U}(t)$ must satisfy the normalization condition $|\alpha (t)|^{2}+|\beta
(t)|^{2}=1$. Under this evolution, a given source state $|s\rangle $, which
can be written as 
\begin{equation}
\left\vert s\right\rangle \overset{\text{def}}{=}x\left\vert w\right\rangle +%
\sqrt{1-x^{2}}\left\vert w_{\perp }\right\rangle \text{,}
\end{equation}%
will be mapped as follows, 
\begin{equation}
\binom{x}{\sqrt{1-x^{2}}}\rightarrow \binom{\alpha \left( t\right) x+\beta
\left( t\right) \sqrt{1-x^{2}}}{-\beta ^{\ast }\left( t\right) x+\alpha
^{\ast }\left( t\right) \sqrt{1-x^{2}}}\text{,}
\end{equation}%
where $x\overset{\text{def}}{=}\langle w|s\rangle $ denotes the quantum
overlap, which can be defined to be real. The transition probability from $%
|s\rangle $ to $|w\rangle $ can then be straightforwardly computed as, 
\begin{equation}
\mathcal{P}_{\left\vert s\right\rangle \rightarrow \left\vert w\right\rangle
}\left( t\right) \overset{\text{def}}{=}\left\vert \left\langle w|\mathcal{U}%
\left( t\right) |s\right\rangle \right\vert ^{2}=\left\vert \alpha \left(
t\right) \right\vert ^{2}x^{2}+\left\vert \beta \left( t\right) \right\vert
^{2}\left( 1-x^{2}\right) +\left[ \alpha \left( t\right) \beta ^{\ast
}\left( t\right) +\alpha ^{\ast }\left( t\right) \beta \left( t\right) %
\right] x\sqrt{1-x^{2}}\text{.}  \label{good}
\end{equation}%
Therefore, one needs to compute the probability amplitudes $\alpha (t)$ and $%
\beta (t)$ from the evolution operator $\mathcal{U}(t)$ in order to find an
exact analytical expression for the transition probability in Eq. (\ref{good}%
). This is in general however, a rather difficult task. In terms of the
fields $\omega (t)$ and $\Omega (t)$ defined previously and using the
relation $i\hslash \,\dot{\mathcal{U}}(t)=\mathcal{H}_{\mathrm{su}(2\text{; }%
\mathbb{C})}(t)\mathcal{U}(t)$, one must solve the following coupled system
of first-order ordinary differential equation with time-dependent
coefficients, 
\begin{equation}
i\hslash \dot{\alpha}\left( t\right) =\Omega \left( t\right) \alpha \left(
t\right) -\omega \left( t\right) \beta ^{\ast }\left( t\right) \text{, and }%
i\hslash \dot{\beta}\left( t\right) =\omega \left( t\right) \alpha ^{\ast
}\left( t\right) +\Omega \left( t\right) \beta \left( t\right) \text{,}
\label{lodes}
\end{equation}%
subject to the initial conditions $\mathcal{U}(t)=\mathcal{I}$, $\alpha
\left( 0\right) =1$ and $\beta \left( 0\right) =0$. Unfortunately, this
approach does not lead to exact analytical solutions in many cases.

It is well known that it is difficult to exactly solve time-dependent
two-level quantum system evolutions like that described above for arbitrary
magnetic field configurations. There are however, special cases that have
been studied and found to have exact analytical solutions. Landau and Zener
studied the case of a spin-$1/2$ particle immersed in a magnetic field of
constant intensity and time-dependent frequency \cite{landau32,zener32a}. In
a subsequent work, Rabi considered a similar scenario involving a
time-dependent magnetic field precessing around the quantized $z$-axis,
except with constant frequency \cite{rabi37,rabi54}. Rosen and Zener have
also investigated the dynamics of an electron moving in an external magnetic
field with a hyperbolic secant pulse configuration, in a purely analytical
manner \cite{zener32}. In these scenarios, precise expressions for the
chosen magnetic fields are justified either by the physics of the system
being studied or (in the case of Rosen and Zener) by the mathematical
convenience leading to an analytically integrable system of coupled
differential equations. For more examples of novel mathematical techniques
leading to systematic approaches to solving problems of this kind, we refer
to Refs. \cite{barnes12,messina14,grimaudo18}. Despite the difficulty of
analytically solving this problem in general, it is occasionally possible to
obtain exact solutions given some convenient physically or mathematically
motivated magnetic field configurations.

One very powerful technique often used in magnetic resonance scenarios is
called the rotating coordinates technique, originally used by Rabi, Ramsey,
and Schwinger. In this technique one transforms the given\textbf{\ }problem
to a convenient coordinate system that rotates with the system of interest,
solves the simplified problem in the new system of coordinates, then applies
the inverse transformation thereby facilitating a return to the original
coordinates. In this context, the new coordinates are frequently referred to
as the \textquotedblleft rotating frame\textquotedblright , while the
original coordinates are often called the \textquotedblleft stationary
frame\textquotedblright . In the case of resonance experiments, the rotating
frame is usually chosen to rotate about the axis defined by the magnetic
field $\vec{B}_{\parallel }$ with angular frequency matching that of $\vec{B}%
_{\perp }$. To be explicit, assume that the quantum mechanical evolution in
the stationary frame \textbf{is} governed by the Schr\"{o}dinger equation, 
\begin{equation}
i\hslash \partial _{t}\left\vert \psi \left( t\right) \right\rangle =%
\mathcal{H}\left( t\right) \left\vert \psi \left( t\right) \right\rangle 
\text{.}
\end{equation}%
If we define a unitary transformation $T$ that maps between two distinct
bases $\{|\psi (t)\rangle \}$ and $\{|\psi ^{\prime }(t)\rangle \}$, then
the evolution in the new frame of reference is governed by 
\begin{equation}
i\hslash \partial _{t}\left\vert \psi ^{\prime }\left( t\right)
\right\rangle =\mathcal{H}^{\prime }\left( t\right) \left\vert \psi ^{\prime
}\left( t\right) \right\rangle \text{,}
\end{equation}%
where $\left\vert \psi ^{\prime }\left( t\right) \right\rangle \overset{%
\text{def}}{=}T\left\vert \psi \left( t\right) \right\rangle $ and $\mathcal{%
H}^{\prime }\left( t\right) \overset{\text{def}}{=}\left[ T\mathcal{H}\left(
t\right) T^{\dagger }+i\hslash \left( \partial _{t}T\right) T^{\dagger }%
\right] $. By applying this result to our generalized Hamiltonian models, we
are able to consider the particular case of a transformation $T\overset{%
\text{def}}{=}e^{-i\sigma _{z}\frac{\phi _{\omega }\left( t\right) }{2}}$
that rotates coordinates by an angle $\phi _{\omega }\left( t\right) $ about
the $z$-axis. Then the Hamiltonians $\mathcal{H}_{\mathrm{su}\left( 2\text{; 
}\mathbb{C}\right) }\left( t\right) $ in Eqs. (\ref{peter2}) and (\ref%
{peter3}) become 
\begin{equation}
\mathcal{H}_{\mathrm{su}\left( 2\text{; }%
\mathbb{C}
\right) }^{\prime }\left( t\right) \overset{\text{def}}{=}\left[ \Omega
\left( t\right) +\frac{\hslash }{2}\dot{\phi}_{\omega }\left( t\right) %
\right] \sigma _{z}+\omega _{\mathcal{H}}\left( t\right) \sigma _{x}\text{,}
\label{rabi1}
\end{equation}%
and,%
\begin{equation}
\mathcal{H}_{\mathrm{su}\left( 2\text{; }%
\mathbb{C}
\right) }^{\prime }\left( t\right) \overset{\text{def}}{=}\left[ -\frac{%
e\hslash }{2mc}B_{\parallel }\left( t\right) +\frac{\hslash }{2}\dot{\phi}%
_{\omega }\left( t\right) \right] \sigma _{z}+\frac{e\hslash }{2mc}B_{\perp
}\left( t\right) \sigma _{x}\text{,}  \label{rabi2}
\end{equation}%
respectively, where $\dot{\phi}_{\omega }\overset{\text{def}}{=}d\phi
_{\omega }/dt$. The classic Rabi scenario \cite{sakurai94} concerns the
special case where the magnetic field intensities $B_{\parallel }$ and $%
B_{\perp }$ are constant while $\phi _{\omega }(t)\overset{\text{def}}{=}%
\omega _{0}t$ with a negative constant $\omega _{0}$. Choosing $\omega _{0}$
based on the so-called static resonance condition, 
\begin{equation}
\omega _{0}=\frac{e}{mc}B_{\parallel }\text{,}
\end{equation}%
the Hamiltonian in Eq. (\ref{rabi2}) becomes time-independent. Therefore,
the Rabi scenario provides an example whereby the rotating coordinate
technique transforms a time-dependent Hamiltonian into a time-independent
quantum mechanical problem. In a generalization of this scenario presented
by Messina and collaborators \cite{messina14,grimaudo18}, the quantities $%
B_{\parallel }$, $B_{\perp }$, and $\phi _{\omega }$ are considered to be
time-dependent dynamical variables. In this case, one can impose a
generalized Rabi condition, 
\begin{equation}
\dot{\phi}_{\omega }\left( t\right) =-\frac{2}{\hslash }\Omega \left(
t\right) \text{,}  \label{GRC}
\end{equation}%
in which case the Hamiltonians in Eqs. (\ref{rabi1}) and (\ref{rabi2}) lead
to a simplified but not time-independent quantum mechanical problem.

As a side remark about experimental considerations, one can understand
typical values of $B_{\parallel }$ and $B_{\perp }$ by noting that readout
in nuclear magnetic resonance (NMR) experiments usually involve measuring
the time-varying magnetization of a substance under the action of an
external magnetic field. Specifically, the behavior of this magnetization
determines the time constants commonly denoted by $T_{1}$ and $T_{2}$ \cite%
{bloch46,bloch46b,bloch53}. The quantity $T_{1}$ is the \textquotedblleft
longitudinal\textquotedblright\ ($z$-axis) relaxation time and represents
the characteristic time for an ensemble of spin-$1/2$ particles to return to
thermal equilibrium about the $z$-axis. The quantity $T_{2}$ is the
\textquotedblleft transverse\textquotedblright\ ($xy$-plane) relaxation time
(also known as coherence time), and represents the characteristic time for
spins to lose coherence about the $z$-axis, or for the $x$ and $y$
components of magnetization to average to zero. Nuclear induction
experiments by Bloch and collaborators \cite{bloch46b} used magnetic field
intensities of approximately $B_{\parallel }\simeq 1826\text{ }\mathrm{G}$
and $B_{\perp }\simeq 5\text{ }\mathrm{G}$, with relaxation times ranging
from fractions of a second to several seconds (for example, $T_{1}\simeq
10^{-5}\text{ }\mathrm{sec.}$). The large range of relaxation times can be
attributed to a variety of factors, such as the electronic structure of the
atoms in the substance as well as their distance, motion, and nuclear
moments \cite{bloch46b}.

In our present investigation, we exploit our previous work in Ref. \cite%
{alsing19} and utilize the results of Messina and collaborators in Refs. 
\cite{messina14, grimaudo18} to facilitate the characterization of four
quantum mechanical scenarios in which transition probabilities\textbf{\ }can
be obtained in an exact analytical manner. We consider the transition
probability $\mathcal{P}_{|w_{\perp }\rangle \rightarrow |w\rangle }(t)$
from an initial state $|w_{\perp }\rangle $ to a final state $|w\rangle $
such that $\langle w_{\perp }|w\rangle =0$ and without loss of generality,
we assume these states to be the eigenstates of the Pauli operator $\sigma
_{z}$ (in particular, $\sigma _{z}|w\rangle =+|w\rangle $ and $\sigma
_{z}|w_{\perp }\rangle =-|w_{\perp }\rangle $). In all four scenarios, we
assume the physical conditions are such that 
\begin{equation}
\dot{\phi}_{\omega }\left( t\right) =\omega _{0}\text{, and }\Omega \left(
t\right) =-\frac{\hslash }{2}\omega _{0}\text{,}  \label{chosen}
\end{equation}%
where $\omega _{0}$ is a negative constant.\textbf{\ }It is worth noting
that we need not have chosen this condition in particular. From a general
perspective, any temporal behavior $\dot{\phi}_{\omega }(t)$ and $\Omega (t)$
that satisfies Eq. (\ref{GRC}) would be sufficient, but we choose Eq. (\ref%
{chosen}) because it appears reasonable from an experimental viewpoint. We
can formally categorize the four scenarios by the time-dependent functional
form of the field intensity $\omega _{\mathcal{H}}(t)$. The first case
assumes a constant field intensity, 
\begin{equation}
\omega _{\mathcal{H}}^{\left( 1\right) }\left( t\right) \overset{\text{def}}{%
=}\Gamma \text{,}  \label{32}
\end{equation}%
which defines the original Rabi condition in which $\mathcal{P}_{\left\vert
w_{\perp }\right\rangle \rightarrow \left\vert w\right\rangle }\left(
t\right) $ takes the form, 
\begin{equation}
\mathcal{P}_{\left\vert w_{\perp }\right\rangle \rightarrow \left\vert
w\right\rangle }^{\left( 1\right) }\left( t\right) =\sin ^{2}\left( \frac{%
\Gamma }{\hslash }t\right) \text{.}  \label{tp1}
\end{equation}%
In the other three cases, we consider generalized Rabi scenarios wherein the
field intensity $\omega _{\mathcal{H}}\left( t\right) $ changes with time. In%
\textbf{\ }particular, we investigate three different time-dependent
behaviors: oscillatory, power law decay, and exponential law decay. The
functional form of the field intensities for these cases are given by, 
\begin{equation}
\omega _{\mathcal{H}}^{\left( 2\right) }\left( t\right) \overset{\text{def}}{%
=}\Gamma \cos \left( \lambda t\right) \text{, }\omega _{\mathcal{H}}^{\left(
3\right) }\left( t\right) \overset{\text{def}}{=}\frac{\Gamma }{\left(
1+\lambda t\right) ^{2}}\text{ and }\omega _{\mathcal{H}}^{\left( 4\right)
}\left( t\right) \overset{\text{def}}{=}\Gamma e^{-\lambda t}\text{,}
\label{33}
\end{equation}%
respectively. We assume $\Gamma $ to be a positive constant. Note that $%
\omega _{\mathcal{H}}^{(2)}(t)$ is positive when $0\leq t\leq \pi /(2\lambda
)$, whereas $\omega _{\mathcal{H}}^{(3)}(t)$ and $\omega _{\mathcal{H}%
}^{(4)}(t)$ are strictly positive $\forall t\in \mathbb{R}$. It can be shown
that in all three of these cases, the transition probability $\mathcal{P}%
_{\left\vert w_{\perp }\right\rangle \rightarrow \left\vert w\right\rangle
}^{\left( j\right) }\left( t\right) $ with $j\in \{2,3,4\}$ is given by \cite%
{grimaudo18},%
\begin{equation}
\mathcal{P}_{\left\vert w_{\perp }\right\rangle \rightarrow \left\vert
w\right\rangle }^{\left( j\right) }\left( t\right) =\sin ^{2}\left[
\int_{0}^{t}\frac{\omega _{\mathcal{H}}^{\left( j\right) }\left( t^{\prime
}\right) }{\hslash }dt^{\prime }\right] \text{.}  \label{tp2}
\end{equation}%
This implies that, under resonance conditions, the transition probability in
all four cases only depends on the time integral of the transverse field
intensity $\omega _{\mathcal{H}}(t)$.

We remark that the chosen expressions of $\omega _{\mathcal{H}}(t)$ serve to
specify particular types of behavior in analog quantum search algorithms.
Recall that in Grover's original search algorithm \cite{grover97}, the
algorithm must end at a specific instant in order\textbf{\ }to succeed with
high probability. An improved version of his algorithm \cite{grover05} has a
fixed-point property that allows the success probability to remain high even
when the algorithm runs with more iterations than necessary. Both of these
algorithms are digital (discrete-time) search algorithms. The first analog
(continuous-time) version of Grover's quantum search algorithm was proposed
by Farhi-Gutmann in Ref. \cite{farhi98}. Inspired by works in Refs. \cite%
{byrnes18,dalzell17}, we provided a perspective on the connections between
analog quantum search algorithms and the physics of two-level quantum
systems in Ref. \cite{alsing19}. We investigated this connection in detail
by means of exactly solvable time-dependent two-level quantum systems,
starting from a generalized quantum search Hamiltonian originally proposed
in Ref. \cite{bae02}. In particular, we analytically computed the transition
probabilities associated with a number of physical scenarios in which a spin-%
$1/2$ particle is immersed in an external time-dependent magnetic field.
Furthermore, we analyzed the temporal behaviors of these transition
probabilities (oscillatory and monotonic) in analogy with key features of
different types of quantum search algorithms (Grover-like and fixed point,
respectively). We then investigated the physical connection between a search
algorithm's schedule and the control fields in a driving Hamiltonian in both
adiabatic \cite{roland02} and nonadiabatic \cite{romanelli07} quantum
mechanical evolutions. The present paper builds upon this line of
investigation by specifying, in terms of the transverse field intensity $%
\omega _{\mathcal{H}}(t)$, the particular behavior associated with analog
quantum search algorithms corresponding to the two-level system.

Table II summarizes the main characteristics of the four quantum mechanical
evolutions that we consider in this paper. In the next section, we will use
the transition probabilities in Eqs. (\ref{tp1}) and (\ref{tp2}) to
construct our parametrized output quantum states.

\begin{table}[t]
\centering
\begin{tabular}{c|c|c|c}
\hline\hline
Rabi Scenario & Transversal Magnetic Field Intensity, $B_{\perp}\left(
t\right) $ & Resonance Condition & Complex Transverse Field, $\omega\left(
t\right) $ \\ \hline
original & $\left( 2mc/\left\vert e\right\vert \hslash\right) \Gamma$ & $%
B_{\parallel}=\left( mc/e\right) \omega_{0}$ & $\Gamma e^{i\phi_{\omega
}\left( t\right) }$ \\ 
generalized & $\left( 2mc/\left\vert e\right\vert \hslash\right) \omega_{%
\mathcal{H}}\left( t\right) $ & $B_{\parallel}\left( t\right) =\left(
mc/e\right) \dot{\phi}_{\omega}\left( t\right) $ & $\Gamma \cos\left(
\lambda t\right) e^{i\phi_{\omega}\left( t\right) }$ \\ 
generalized & $\left( 2mc/\left\vert e\right\vert \hslash\right) \omega_{%
\mathcal{H}}\left( t\right) $ & $B_{\parallel}\left( t\right) =\left(
mc/e\right) \dot{\phi}_{\omega}\left( t\right) $ & $\Gamma\left( 1+\lambda
t\right) ^{-2}e^{i\phi_{\omega}\left( t\right) }$ \\ 
generalized & $\left( 2mc/\left\vert e\right\vert \hslash\right) \omega_{%
\mathcal{H}}\left( t\right) $ & $B_{\parallel}\left( t\right) =\left(
mc/e\right) \dot{\phi}_{\omega}\left( t\right) $ & $\Gamma e^{-\lambda
t}e^{i\phi_{\omega}\left( t\right) }$ \\ \hline
\end{tabular}%
\caption{Illustrative representation of the complex transverse field $%
\protect\omega\left( t\right) $, the resonance condition in terms of the
longitudinal magnetic field intensity $B_{\parallel}\left( t\right) $, and
the transverse magnetic field intensity $B_{\perp}\left( t\right) $ in the
selected four Rabi scenarios.}
\end{table}

\section{Optimum paths, entropic speed, and entropy production rate}

In this section, we use the Fisher information (see Eq. (\ref{JIJ2}) in
Section II) to introduce a metric on a manifold of probability vectors,
where the latter represents\textbf{\ }transition probabilities in each of
the four quantum mechanical evolutions generated by our generalized $\mathrm{%
su}\left( 2\text{; }\mathbb{C}\right) $ Hamiltonian models (see Eqs. (\ref%
{tp1}) and (\ref{tp2}) in Section III). By using a minimum-action principle
to deduce the optimum path on this manifold between an initial source state
and a final target state, we verify that this optimum path is the geodesic
(shortest) path between the two points representing the states and further,
this path also serves to minimize the total entropy production during the
evolution. We also demonstrate by explicitly calculating the entropic speed
(see Eq. (\ref{vthermo})) and the entropy production rate (see Eq. (\ref{EPR}%
) in each scenario, that to a faster transition between states there
necessarily corresponds a higher rate of entropy production. For an
intriguing presentation on the link between the geometry of evolving
probabilities and quantum theory, we refer to Ref. \cite{reginatto}.

\subsection{Use of the Fisher information in physics}

To motivate our analysis, we note that the Fisher information has been used
to characterize the evolution of many different kinds of physical systems.
It was first introduced by Fisher \cite{fisher25} in statistical estimation
theory, and was utilized by Linnik \cite{linnik59} as part of an
information-theoretic proof of the central limit theorem in probability
theory. For this reason, the Fisher information is sometimes called the
Linnik functional. However, it has also been used to prove results of the
kinetic theory of gases. It was noted by McKean that the Fisher information
monotonically decreases with time along solutions to Kac's model of a
one-dimensional Maxwellian gas \cite{mckean66}, and this result was later
extended to gases in two-dimensional \cite{toscani92} and any-dimensional 
\cite{villani98a} velocity spaces by Toscani and Villani, respectively.
Villani also proved the same monotonically decreasing behavior of the Fisher
information along solutions to the spatially homogeneous Landau equation,
again for a Maxwellian gas in any dimension of velocity space \cite%
{villani98b,villani00}.

Inspired by these ideas, our information geometric characterization of paths
in quantum mechanical evolution focuses on the Fisher information and its
behavior along solutions to a differential equation encoding the constraint
of minimum entropy production.

\subsection{Deriving probability paths from quantum evolution}

Recall that in Section II (see Eq. (\ref{fubini2})) we exploited the phase
indeterminacy of state vectors in quantum mechanics to make the Fubini-Study
metric tensor components $g_{\alpha \beta }^{\text{(FS)}}(\theta )$
proportional to the Fisher metric tensor components $\mathcal{F}_{\alpha
\beta }(\theta )$. In particular, we chose a basis of orthonormal vectors $%
\{|w\rangle ,|w_{\perp }\rangle \}$ for the $n$-qubit Hilbert space $%
\mathcal{H}_{2}^{n}$ such that $\mbox{Im}\left( \left\langle \psi
|w\right\rangle \left\langle w|d\psi _{\perp }\right\rangle \right) =0=%
\mbox{Im}\left( \left\langle \psi |w_{\perp }\right\rangle \left\langle
w_{\perp }|d\psi _{\perp }\right\rangle \right) $, which caused the
non-negative variance of the phase changes to vanish in the definition of $%
g_{\alpha \beta }^{\text{(FS)}}(\theta )$ \cite{caves94, cafaropre18}.
Recall also that in Section III, we derived transition probabilities $%
\mathcal{P}_{|w_{\perp }\rangle \rightarrow |w\rangle }(t)$ for four special
cases starting from a generalized Hamiltonian for a two-level quantum
system. What we now consider, building upon the connections explored between
analog quantum search algorithms and two-level quantum systems \cite%
{alsing19,alsing19b}, is a set of output states $|\psi (\theta )\rangle $ of
a continuous-time quantum search algorithm. Expressed in our basis, this
takes the form, 
\begin{equation}
\left\vert \psi \left( \theta \right) \right\rangle \overset{\text{def}}{=}%
e^{i\varphi _{w}\left( \theta \right) }\sqrt{p_{w}\left( \theta \right) }%
\left\vert w\right\rangle +e^{i\varphi _{w_{\perp }}\left( \theta \right) }%
\sqrt{p_{w_{\perp }}\left( \theta \right) }\left\vert w_{\perp
}\right\rangle \text{.}  \label{output}
\end{equation}%
We assume the input to the algorithm to be an $N=2^{n}$-dimensional
normalized $n$-qubit source state $|s\rangle \overset{\text{def}}{=}|\psi
(\theta _{0})\rangle $, such that $|\psi (\theta )\rangle \in \mathcal{H}%
_{2}^{n}$ and the $n$-qubit Hilbert space is spanned by $\{|w\rangle
,|w_{\perp }\rangle \}$. We interpret the parameter $\theta $ as the elapsed
time $t$ in a statistical sense. More specifically, $\theta $ represents
some measurable observable varying with time that can be determined
experimentally (for example, a time-varying magnetic field intensity). We
note that treating elapsed time as an experimentally controllable
statistical parameter is not an unusual idea. Indeed, it is not dissimilar
from the idea of the Wick rotation \cite{wick54}, in which the inverse
temperature $\beta \overset{\text{def}}{=}(k_{B}T)^{-1}$ (where $k_{B}\simeq
1.38\times 10^{-23}$ $\left[ \mathrm{MKSA}\right] $ is the Boltzmann
constant) is replaced with the imaginary time $it/\hslash $ so as to turn
calculations of quantum mechanical transition amplitudes into calculations
resembling statistical mechanical averages of observables. One can think of $%
\theta $ as the parameter along which one drives the evolution of the
quantum mechanical system.

The central question can then be stated as\textbf{:} which paths through the
space of states parametrized by $\theta $ are optimal under the constraint
of minimum entropy production? We answer this question by considering the
manifold of probability distributions $p(\theta )$ generated by the output
states $|\psi (\theta )\rangle $ as follows, 
\begin{equation}
\left\vert \psi \left( \theta \right) \right\rangle \mapsto p\left( \theta
\right) =\left( p_{w}\left( \theta \right) \text{, }p_{w_{\perp }}\left(
\theta \right) \right) =\left( \left\vert \left\langle w|\psi \left( \theta
\right) \right\rangle \right\vert ^{2}\text{, }\left\vert \left\langle
w_{\perp }|\psi \left( \theta \right) \right\rangle \right\vert ^{2}\right) 
\text{,}  \label{output2}
\end{equation}%
where $p_{w}\left( \theta \right) \overset{\text{def}}{=}\left\vert
\left\langle w|\psi \left( \theta \right) \right\rangle \right\vert ^{2}$
and $p_{w_{\perp }}\left( \theta \right) \overset{\text{def}}{=}\left\vert
\left\langle w_{\perp }|\psi \left( \theta \right) \right\rangle \right\vert
^{2}$ are the squared probability amplitudes denoting the success and
failure probabilities respectively, of the analog quantum search algorithm.
In our following analysis, we use the Fisher metric tensor in Eq. (\ref{JIJ2}%
) (which, given our working assumptions, is proportional to the Fubini-Study
metric in Eq. (\ref{fubini2})), and without loss of generality, we assume
the initial input state to be $|\psi (\theta _{0})\rangle =|w_{\perp
}\rangle $.

As an illustration of what follows, we show in Fig. $1$ the behavior of the
Fisher information evaluated along the probabilities obtained from the $%
\mathrm{su}\left( 2\text{; }%
\mathbb{C}
\right) $ quantum evolution with an exponentially decaying transverse field
intensity, given by $\omega _{\mathcal{H}}^{\left( 4\right) }\left( t\right) 
$ in Eq. (\ref{33}).

\begin{figure}[t]
\centering
\includegraphics[width=1\textwidth] {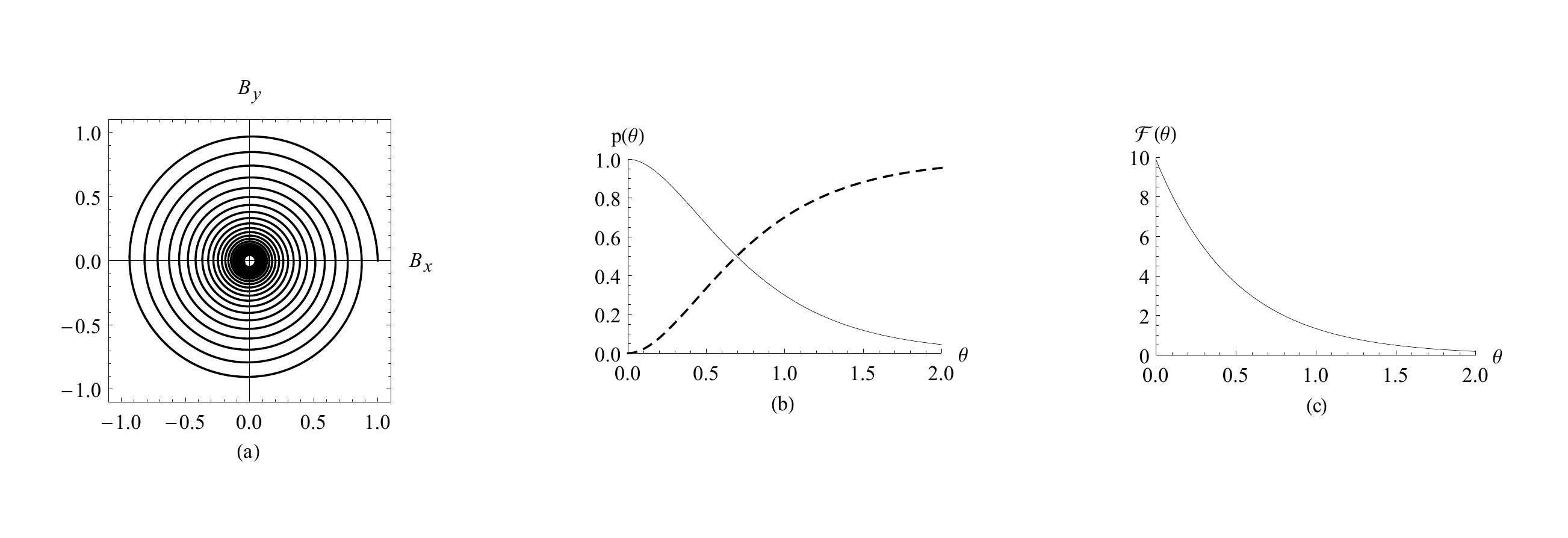}
\caption{In plot (c), we depict the behavior of the Fisher information $%
\mathcal{F}\left( \protect\theta \right) $ versus $\protect\theta $. The
Fisher information is defined in terms of the transition probabilities $%
p_{w}\left( \protect\theta \right) $ (dashed line) and $p_{w_{\bot }}\left( 
\protect\theta \right) $ (solid line) that appear in plot (b). These
probabilities, in turn, arise from the particular external magnetic field $%
\vec{B}=\vec{B}_{\bot }+\vec{B}_{\parallel }$ with $\vec{B}_{\bot }=B_{x}%
\hat{x}+B_{y}\hat{y}$ depicted in plot (a) that specifies the $\mathrm{su}%
\left( 2\text{; }\mathbb{C}\right) $ Hamiltonian model being studied. In
plots (b) and (c), we set $\protect\lambda =1$ and $\Gamma /\hslash \protect%
\lambda =\protect\pi /2$ and. In plot (a), the two-dimensional parametric
plot of the transverse magnetic field components, we set $\protect\lambda =1$
and $\left\vert \protect\omega _{0}\right\vert =15\protect\pi $.
Furthermore, for simplifying normalization purposes, we also set $\Gamma /%
\protect\mu _{\text{Bohr}}=1$ where $\protect\mu _{\text{Bohr}}$ denotes the
Bohr magneton in plot (a). All physical quantities are assumed to be
conveniently expressed by means of the MKSA unit system. Finally, the $%
\mathrm{su}\left( 2\text{; }\mathbb{C}\right) $ Hamiltonian model taken into
consideration in this figure corresponds to the fourth scenario analyzed in
this manuscript.}
\label{fig1}
\end{figure}

\subsection{Illustrative examples: Four types of driving scheme}

The form of the Fisher information $\mathcal{F}(\theta )$ as a function of
the driving parameter $\theta $ is fundamental to the structure of the
manifold, and therefore to the behavior of the paths $\gamma _{\theta }$
with $\theta =\theta (\xi )$ (where $\xi $ is an affine parameter) which
describes the evolution.

For the sake of clarity, we point out that optimum cooling paths are
Riemannian geodesic paths with respect to the dimensionless thermodynamic
time $t_{\text{\textrm{th}}}$ with $dt_{\text{\textrm{th}}}\overset{\text{def%
}}{=}d\tau ^{\prime }/\tau _{\ast }$ where $\tau ^{\prime }\in \left[ 0\text{%
, }\bar{\tau}\right] $ is the clock-time, while $\bar{\tau}$ and $\tau
_{\ast }$ denotes the duration and the mean internal relaxation time of the
physical process being considered, respectively. With this notation in mind,
the divergence in Eq. (\ref{divergence}) can be recast as \cite%
{salamon83,diosi96},%
\begin{equation}
\mathcal{I}\left( \bar{\tau}\right) \overset{\text{def}}{=}\tau _{\ast
}\int_{0}^{\bar{\tau}}\frac{d\theta ^{\alpha }}{d\tau ^{\prime }}g_{\alpha
\beta }\left( \theta \right) \frac{d\theta ^{\beta }}{d\tau ^{\prime }}d\tau
^{\prime }=\int_{0}^{\bar{\tau}/\tau _{\ast }}\frac{d\theta ^{\alpha }}{dt_{%
\text{\textrm{th}}}}g_{\alpha \beta }\left( \theta \right) \frac{d\theta
^{\beta }}{dt_{\text{\textrm{th}}}}dt_{\text{\textrm{th}}}\text{.}
\end{equation}%
Upon identifying the duration of the process $\tau $ with $\bar{\tau}/\tau
_{\ast }$ and the affine parameter $\xi $ with the the dimensionless
thermodynamic time $t_{\text{\textrm{th}}}$,\textbf{\ }the divergence can
henceforth\textbf{\ }be written as%
\begin{equation}
\mathcal{I}\left( \tau \right) \overset{\text{def}}{=}\int_{0}^{\tau }\frac{%
d\theta ^{\alpha }}{d\xi }g_{\alpha \beta }\left( \theta \right) \frac{%
d\theta ^{\beta }}{d\xi }d\xi \text{.}  \label{divergence2}
\end{equation}

In Section III, we introduced four different schemes characterized by
different functional forms of the transverse field intensity $\omega _{%
\mathcal{H}}(t)$ in the generalized Rabi scenario (constant, oscillatory,
power-law decay, and exponential decay). The transition probabilities in
each of these schemes (see Eq. (\ref{tp2})) will generate a corresponding
functional form for the Fisher information, which in turn determines the
form of the geodesic equations which describe the optimum paths. One can
then compute from these optimum paths a characteristic entropic speed $v_{E}$
and entropy production rate $r_{E}$ that differs in each of the four cases.

\subsubsection{Constant Fisher information}

The first case we consider is the case of constant Fisher information,
generated by a constant temporal behavior of the transverse field intensity
(see Eqs. (\ref{32}) and (\ref{tp1})). Therefore, as specified in Eqs. (\ref%
{output}) and (\ref{output2}), we consider the manifold of probability
distributions $\left\{ p\left( \theta \right) \right\} $ with $p\left(
\theta \right) \overset{\text{def}}{=}\left( p_{w}\left( \theta \right) 
\text{, }p_{w_{\perp }}\left( \theta \right) \right) $ where the success and
failure probabilities are given by,%
\begin{equation}
p_{w}\left( \theta \right) \overset{\text{def}}{=}\sin ^{2}\left( \frac{%
\Gamma }{\hslash }\theta \right) \text{, and }p_{w_{\perp }}\left( \theta
\right) \overset{\text{def}}{=}\cos ^{2}\left( \frac{\Gamma }{\hslash }%
\theta \right) \text{,}  \label{1a}
\end{equation}%
respectively. In this case, the probabilities oscillate with a period of $T%
\overset{\text{def}}{=}\pi \hslash /\Gamma $. The Fisher information (Eq. (%
\ref{JIJ2})) is a constant, $\mathcal{F}_{0}$, given by, 
\begin{equation}
\mathcal{F}\left( \theta \right) =\mathcal{F}_{0}\overset{\text{def}}{=}%
4\left( \frac{\Gamma }{\hslash }\right) ^{2}\text{.}
\end{equation}%
Since we are considering a single-parameter probability space, the geodesic
equations in Eq. (\ref{general}) simplify to, 
\begin{equation}
\frac{d^{2}\theta }{d\xi ^{2}}+\frac{1}{2\mathcal{F}}\frac{d\mathcal{F}}{%
d\theta }\left( \frac{d\theta }{d\xi }\right) ^{2}=0\text{.}  \label{odeti}
\end{equation}%
Moreover, since $\frac{d\mathcal{F}}{d\theta }=0$, Eq. (\ref{odeti}) is
equivalent to, 
\begin{equation}
\frac{d^{2}\theta }{d\xi ^{2}}=0\text{,}  \label{geo1}
\end{equation}%
which, assuming initial conditions $\theta (\xi _{0})=\theta _{0}\in \mathbb{%
R}_{+}$ and $\dot{\theta}(\xi _{0})=\dot{\theta}_{0}\in \mathbb{R}_{+}$,
yields simple optimum paths of the form, 
\begin{equation}
\theta \left( \xi \right) =\theta _{0}+\dot{\theta}_{0}\left( \xi -\xi
_{0}\right) \text{.}  \label{geo1a}
\end{equation}%
We can now compute the entropic speed and the entropy production rate for
these optimum paths. Using Eqs. (\ref{vthermo}) and (\ref{geo1a}), we find
that the entropic speed $v_{\mathrm{E}}$ is given by, 
\begin{equation}
v_{\mathrm{E}}\left( \Gamma \right) =\frac{\Gamma }{\hslash }\dot{\theta}_{0}%
\text{.}  \label{speed1}
\end{equation}%
Furthermore, using Eqs. (\ref{divergence2}) and (\ref{geo1a}), we find that
the rate of total entropy production $r_{\mathrm{E}}$ is given by, 
\begin{equation}
r_{\mathrm{E}}\left( \Gamma \right) =\left( \frac{\Gamma }{\hslash }\right)
^{2}\dot{\theta}_{0}^{2}\text{.}  \label{lambda1}
\end{equation}%
We remark that both the entropic speed and the entropy production rate
depend on the magnitude $\Gamma $ of the transverse field $\omega _{\mathcal{%
H}}^{(1)}(t)$, with $v_{\mathrm{E}}\left( \Gamma \right) \propto \Gamma $
and $r_{\mathrm{E}}\left( \Gamma \right) \propto \Gamma ^{2}$. Therefore,
one can simply manipulate the quantity $\Gamma $ to find a desired\textbf{\ }%
trade-off between speed and entropy production rate in the quantum
mechanical evolution under investigation. This is the first example of such
a trade-off in our information geometric analysis.

\begin{figure}[t]
\centering
\includegraphics[width=0.45\textwidth] {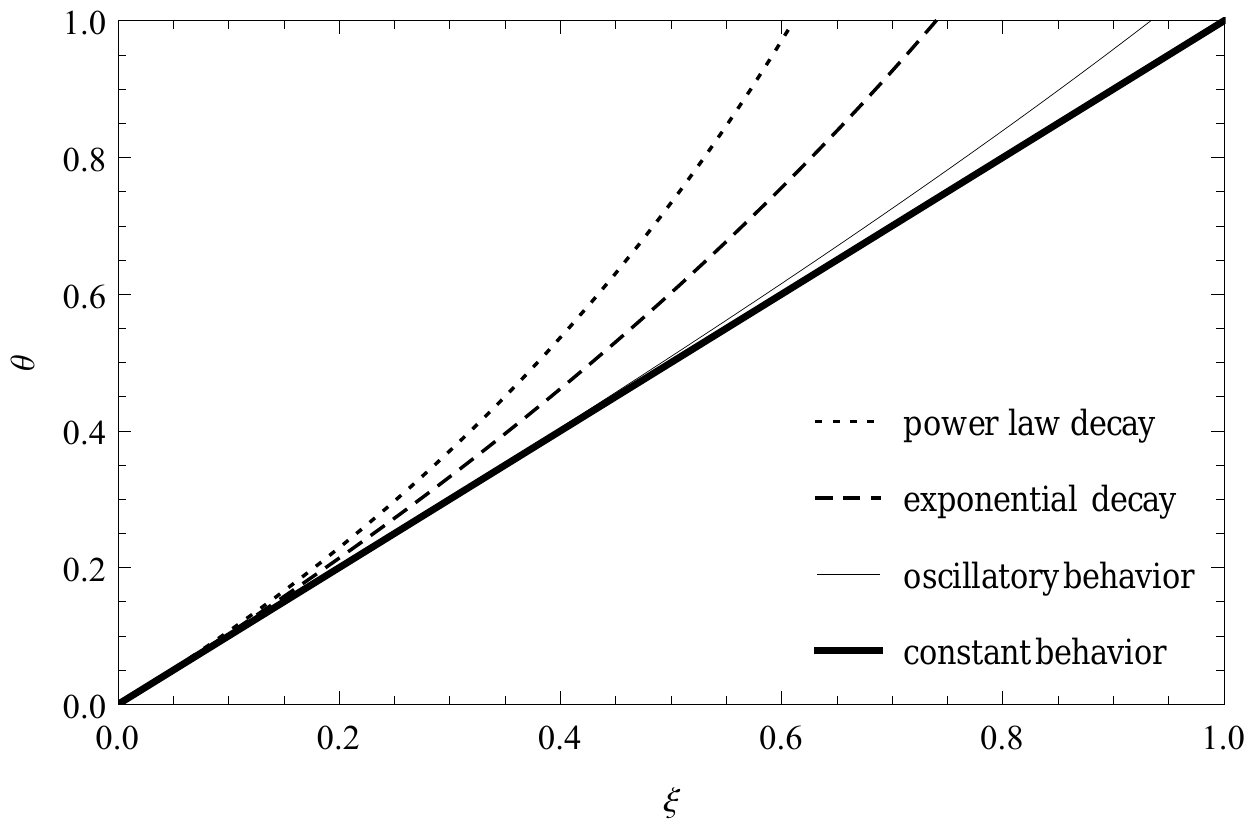}
\caption{Plot of the analytical expressions of the geodesic paths in the
four scenarios being considered: Power law decay of the Fisher information
(dotted), exponential decay of the Fisher information (dashed), oscillatory
behavior of the Fisher information (thin solid), and constant behavior of
the Fisher information (thick solid): In all plots, we set $\protect\theta%
_{0}=0$, $\dot{\protect\theta}_{0}=1$, $\Gamma/\hslash=1$, and $\protect%
\lambda=2/\protect\pi$.}
\label{fig2}
\end{figure}

\subsubsection{Oscillatory behavior of the Fisher information}

The second case we consider is the case of oscillatory Fisher information,
generated by an oscillatory behavior of the transverse field intensity (see
Eqs. (\ref{33}) and (\ref{tp2})). Thus, by recalling Eqs. (\ref{output}) and
(\ref{output2}), we are led to consider the manifold of probability
distributions $\left\{ p\left( \theta \right) \right\} $ with $p\left(
\theta \right) \overset{\text{def}}{=}\left( p_{w}\left( \theta \right) 
\text{, }p_{w_{\perp }}\left( \theta \right) \right) $ where the success and
failure probabilities are of the form, 
\begin{equation}
p_{w}\left( \theta \right) \overset{\text{def}}{=}\sin ^{2}\left[ \frac{%
\Gamma }{\hslash \lambda }\sin \left( \lambda \theta \right) \right] \text{,
and }p_{w_{\perp }}\left( \theta \right) \overset{\text{def}}{=}\cos ^{2}%
\left[ \frac{\Gamma }{\hslash \lambda }\sin \left( \lambda \theta \right) %
\right] \text{,}  \label{5b}
\end{equation}%
respectively. Note that the success probabilities oscillate with a period of 
$T\overset{\text{def}}{=}\pi /\lambda $. For the sake of simplicity, we
shall fix $\Gamma =(h/4)\lambda $ such that $p_{w}(\theta )$ reaches a
maximum value of one at $\theta ^{\ast }\overset{\text{def}}{=}\pi
/(2\lambda )$. We remark that there does not need to be a strict connection
between $\Gamma $ and $\lambda $ in order to achieve a maximum value of one.
In fact, for $\Gamma \geq (h/4)\lambda $, the smallest positive $\theta $
for which $p_{w}(\theta )$ reaches a maximum value of one is given by $%
\tilde{\theta}^{\ast }\overset{\text{def}}{=}(1/\lambda )\sin ^{-1}(\hslash
\lambda /\Gamma )$. By using these expressions for $p_{w}(\theta )$ and $%
p_{w_{\perp }}(\theta )$ to compute the Fisher information, we obtain 
\begin{equation}
\mathcal{F}\left( \theta \right) =4\left( \frac{\Gamma }{\hslash }\right)
^{2}\cos ^{2}\left( \lambda \theta \right) \text{.}  \label{chiappa}
\end{equation}%
With this expression for $\mathcal{F}(\theta )$, the geodesic equation in
Eq. (\ref{odeti}) becomes, 
\begin{equation}
\frac{d^{2}\theta }{d\xi ^{2}}-\lambda \tan \left( \lambda \theta \right)
\left( \frac{d\theta }{d\xi }\right) ^{2}=0\text{.}  \label{difficile}
\end{equation}%
As before, we assume initial conditions $\theta (\xi _{0})=\theta _{0}\in 
\mathbb{R}_{+}$ and $\dot{\theta}(\xi _{0})=\dot{\theta}_{0}\in \mathbb{R}%
_{+}$, and determine that Eq. (\ref{difficile}) yields optimum paths $\theta
(\xi )$ of the following general form, 
\begin{equation}
\theta \left( \xi \right) =\theta _{0}+\frac{\sqrt{1-\lambda ^{2}\xi _{0}^{2}%
}}{\lambda }\dot{\theta}_{0}\left[ \sin ^{-1}\left( \lambda \xi \right)
-\sin ^{-1}\left( \lambda \xi _{0}\right) \right] \text{.}  \label{geo2}
\end{equation}%
These paths specify a particular motion on the statistical manifold, with an
entropic speed $v_{\mathrm{E}}$ given by, 
\begin{equation}
v_{\mathrm{E}}\left( \Gamma \right) =\frac{\Gamma }{\hslash }\left\vert \cos
\left( \lambda \theta _{0}\right) \right\vert \dot{\theta}_{0}\text{,}
\label{speed1b}
\end{equation}%
and a rate of entropy production $r_{\mathrm{E}}$ given by, 
\begin{equation}
r_{\mathrm{E}}\left( \Gamma \right) =\left( \frac{\Gamma }{\hslash }\right)
^{2}\cos ^{2}\left( \lambda \theta _{0}\right) \dot{\theta}_{0}^{2}\text{.}
\label{lambda1b}
\end{equation}%
Compared to the first scenario (Eqs. (\ref{speed1b}) and (\ref{lambda1b}),
this second scenario exhibits geodesic motion characterized by cooler paths
(lower entropy production rate) traversed with slower speed (lower entropic
speed).

\begin{figure}[t]
\centering
\includegraphics[width=1\textwidth] {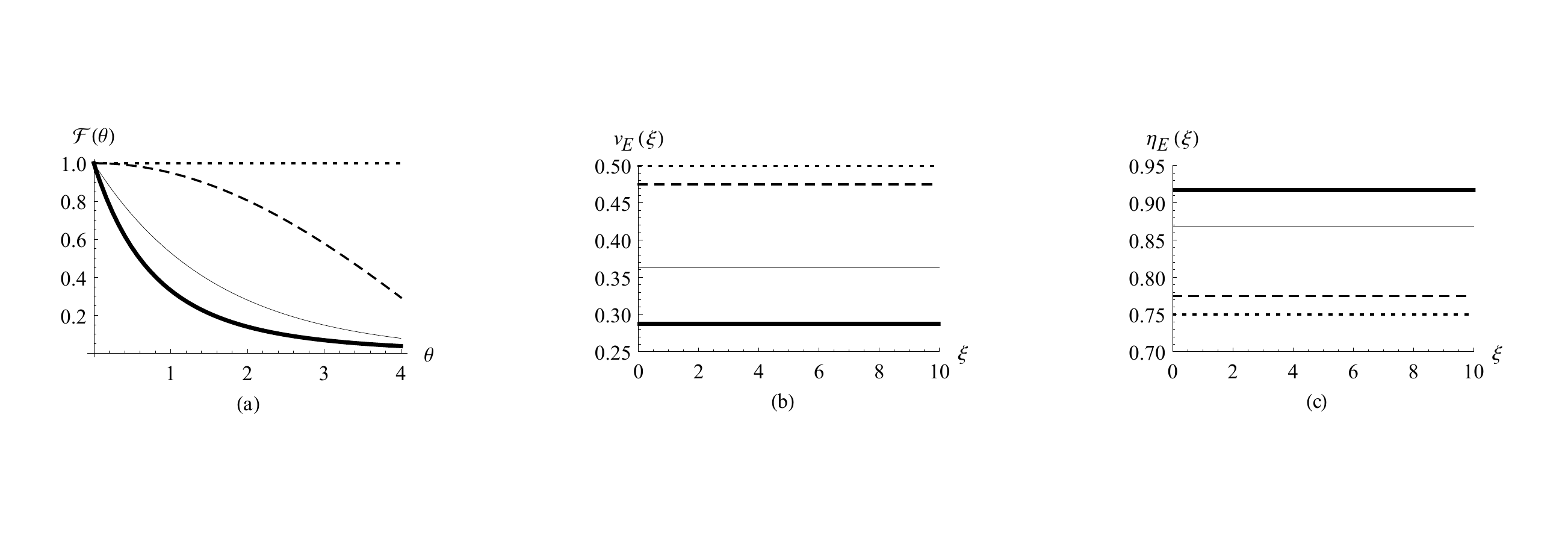}
\caption{In plots (a), (b), and (c), we depict the behavior of the Fisher
information $\mathcal{F}\left( \protect\theta\right) $ versus $\protect%
\theta $, the entropic speed $v_{E}\left( \protect\xi\right) $ versus $%
\protect\xi$, and the entropic efficiency $\protect\eta_{E}\left( \protect\xi%
\right) $ versus $\protect\xi$, respectively. Dotted, dashed, thin solid,
and thick solid lines in plots (a), (b), and (c) correspond to constant,
oscillatory, exponential decay, and power law decay of the Fisher
information, respectively. In all plots, we set $\protect\lambda=1/\protect%
\pi$, $\Gamma/(\hslash\protect\lambda)=\protect\pi/2$, , $\protect\theta %
_{0}=1$, and $\dot{\protect\theta}_{0}=1$. Finally, all physical quantities
are assumed to be conveniently expressed by means of the MKSA unit system.}
\label{fig3}
\end{figure}

\subsubsection{Power law decay of the Fisher information}

The third case we consider is the case of a power-law decay of the Fisher
information, resulting from a power-law behavior of the transverse field
intensity (see Eqs. (\ref{33}) and (\ref{tp2})). Once again exploiting Eqs. (%
\ref{output}) and (\ref{output2}), we consider the manifold of probability
distributions $\left\{ p\left( \theta \right) \right\} $ with $p\left(
\theta \right) \overset{\text{def}}{=}\left( p_{w}\left( \theta \right) 
\text{, }p_{w_{\perp }}\left( \theta \right) \right) $ where the success and
failure probabilities are defined as, 
\begin{equation}
p_{w}\left( \theta \right) \overset{\text{def}}{=}\sin ^{2}\left[ \frac{%
\Gamma }{\hslash \lambda }\left( 1-\frac{1}{1+\lambda \theta }\right) \right]
\text{, and }p_{w_{\perp }}\left( \theta \right) \overset{\text{def}}{=}\cos
^{2}\left[ \frac{\Gamma }{\hslash \lambda }\left( 1-\frac{1}{1+\lambda
\theta }\right) \right] \text{,}  \label{4b}
\end{equation}%
respectively. Observe that when $\Gamma =(h/4)\lambda $, the success
probability $p_{w}(\theta )$ approaches one asymptotically and monotonically
as $\theta $ goes to infinity. With these probabilities, the Fisher
information $\mathcal{F}(\theta )$ takes the form, 
\begin{equation}
\mathcal{F}\left( \theta \right) =4\left( \frac{\Gamma }{\hslash }\right)
^{2}\frac{1}{\left( 1+\lambda \theta \right) ^{4}}\text{,}  \label{rosso}
\end{equation}%
which implies that the geodesic equation in Eq. (\ref{odeti}) becomes, 
\begin{equation}
\frac{d^{2}\theta }{d\xi ^{2}}-\frac{2\lambda }{1+\lambda \theta }\left( 
\frac{d\theta }{d\xi }\right) ^{2}=0\text{.}  \label{domyjob}
\end{equation}%
Once again assuming initial conditions $\theta (\xi _{0})=\theta _{0}\in 
\mathbb{R}_{+}$ and $\dot{\theta}(\xi _{0})=\dot{\theta}_{0}\in \mathbb{R}%
_{+}$, we obtain the following general form for the optimum paths that solve
Eq. (\ref{domyjob}), 
\begin{equation}
\theta \left( \xi \right) =\frac{\left( 1+\lambda \theta _{0}\right)
^{2}+\lambda \dot{\theta}_{0}\left[ \left( \xi -\xi _{0}\right) -\frac{%
1+\lambda \theta _{0}}{\lambda \dot{\theta}_{0}}\right] }{\lambda ^{2}\dot{%
\theta}_{0}\left[ \frac{1+\lambda \theta _{0}}{\lambda \dot{\theta}_{0}}%
-\left( \xi -\xi _{0}\right) \right] }\text{.}  \label{geo3}
\end{equation}%
These paths specify a motion on the statistical manifold with an entropic
speed $v_{\mathrm{E}}$ given by, 
\begin{equation}
v_{\mathrm{E}}\left( \Gamma \right) =\frac{\Gamma }{\hslash }\frac{1}{\left[
1+\lambda \left( \Gamma \right) \theta _{0}\right] ^{2}}\dot{\theta}_{0}%
\text{,}  \label{speed3}
\end{equation}%
and an entropy production rate $r_{\mathrm{E}}$ given by, 
\begin{equation}
r_{\mathrm{E}}\left( \Gamma \right) =\left( \frac{\Gamma }{\hslash }\right)
^{2}\frac{1}{\left[ 1+\lambda \left( \Gamma \right) \theta _{0}\right] ^{4}}%
\dot{\theta}_{0}^{2}\text{.}  \label{lambda3}
\end{equation}%
where we define $\lambda \left( \Gamma \right) \overset{\text{def}}{=}\left(
4\Gamma \right) /h$. We find once again, comparing with the first and second
scenarios, that the present scenario produces even cooler paths with even
slower speed (less entropy production rate and less entropic speed).

\subsubsection{Exponential decay of the Fisher information}

\begin{figure}[t]
\centering
\includegraphics[width=0.45\textwidth] {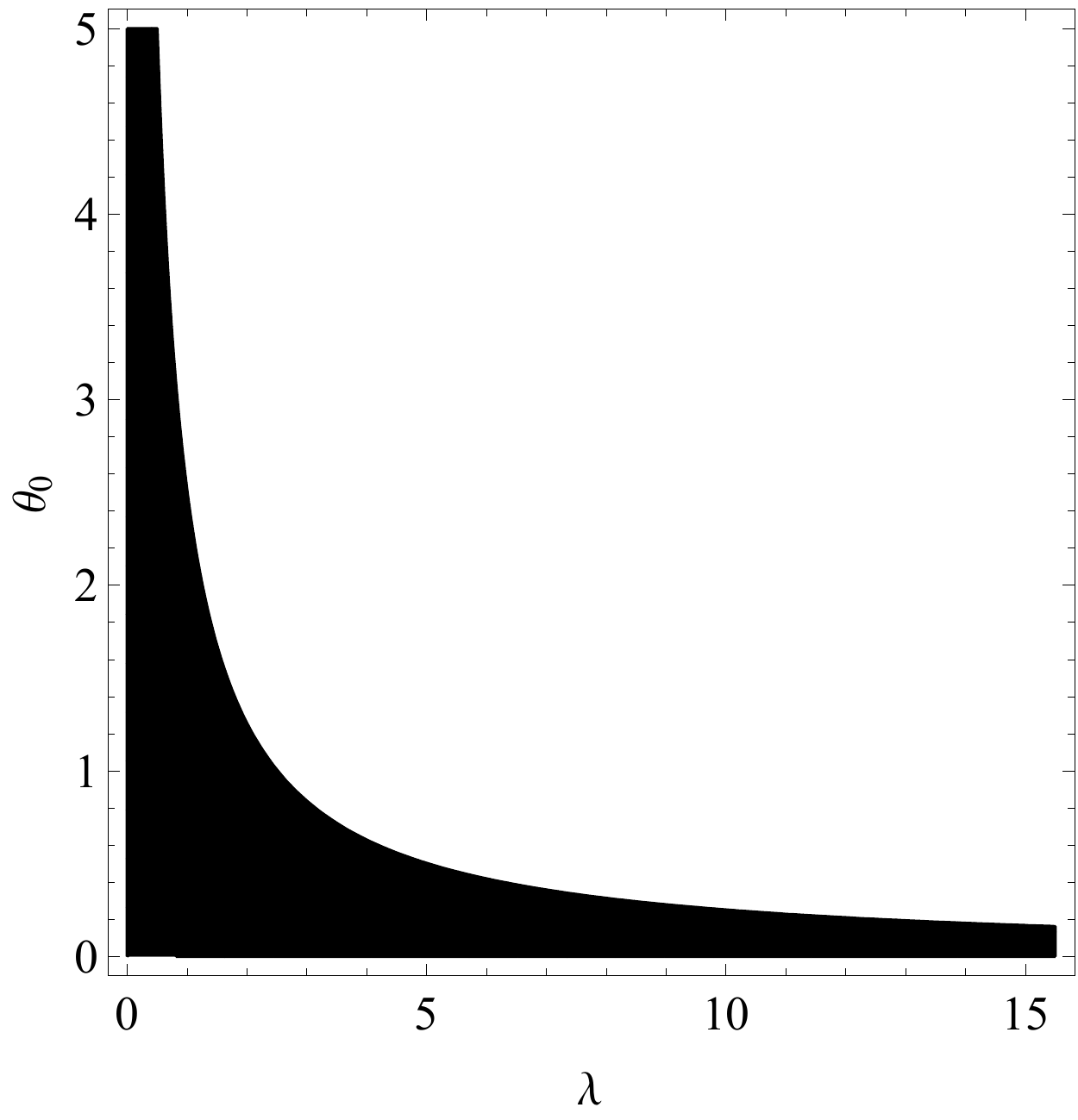}
\caption{Plot of the two-dimensional parametric region $\mathcal{P}$ where
the power law decay strategy is being outperformed by the exponential decay
strategy in terms of higher entropic speed values. As the numerical values
of $\protect\lambda$ become sufficiently large, the black region $\mathcal{P}
$ tends to vanish.}
\label{fig4}
\end{figure}

The fourth and final case we consider is the case of exponentially decaying
Fisher information, produced by an exponentially decaying magnitude of the
transverse field intensity (see Eqs. (\ref{33}) and (\ref{tp2})). Employing
Eqs. (\ref{output}) and (\ref{output2}), we analyze the manifold of
probability distributions $\left\{ p\left( \theta \right) \right\} $ with $%
p\left( \theta \right) \overset{\text{def}}{=}\left( p_{w}\left( \theta
\right) \text{, }p_{w_{\perp }}\left( \theta \right) \right) $ with success
and failure probabilities 
\begin{equation}
p_{w}\left( \theta \right) \overset{\text{def}}{=}\sin ^{2}\left[ \frac{%
\Gamma }{\hslash \lambda }\left( 1-e^{-\lambda \theta }\right) \right] \text{%
, and }p_{w_{\perp }}\left( \theta \right) \overset{\text{def}}{=}\cos ^{2}%
\left[ \frac{\Gamma }{\hslash \lambda }\left( 1-e^{-\lambda \theta }\right) %
\right] \text{,}  \label{3a}
\end{equation}%
respectively. Similarly to the third scenario, letting $\Gamma =(h/4)\lambda 
$ causes the success probability to asymptotically and monotonically
approach the ideal value of one as $\theta $ approaches infinity. The Fisher
information $\mathcal{F}(\theta )$ in this case becomes, 
\begin{equation}
\mathcal{F}\left( \theta \right) =4\left( \frac{\Gamma }{\hslash }\right)
^{2}e^{-2\lambda \theta }\text{,}  \label{verde}
\end{equation}%
which implies that the geodesic equation in Eq. (\ref{odeti}) reduces to, 
\begin{equation}
\frac{d^{2}\theta }{d\xi ^{2}}-\lambda \left( \frac{d\theta }{d\xi }\right)
^{2}=0\text{.}  \label{geq4}
\end{equation}%
Using the same initial conditions $\theta (\xi _{0})=\theta _{0}\in \mathbb{R%
}_{+}$ and $\dot{\theta}(\xi _{0})=\dot{\theta}_{0}\in \mathbb{R}_{+}$ as
before, this geodesic equation yields the following general form for optimum
paths, 
\begin{equation}
\theta \left( \xi \right) =\theta _{0}-\frac{1}{\lambda }\log \left[
1-\lambda \dot{\theta}_{0}\left( \xi -\xi _{0}\right) \right] \text{.}
\label{geo4}
\end{equation}%
The motion characterized by these paths exhibits an entropic speed $v_{%
\mathrm{E}}$ given by, 
\begin{equation}
v_{\mathrm{E}}\left( \Gamma \right) =\frac{\Gamma }{\hslash }e^{-\lambda
\left( \Gamma \right) \theta _{0}}\dot{\theta}_{0}\text{,}  \label{speed4}
\end{equation}%
and a rate of entropy production $r_{\mathrm{E}}$ given by, 
\begin{equation}
r_{\mathrm{E}}\left( \Gamma \right) =\left( \frac{\Gamma }{\hslash }\right)
^{2}e^{-2\lambda \left( \Gamma \right) \theta _{0}}\dot{\theta}_{0}^{2}\text{%
,}  \label{lambda4}
\end{equation}%
where as before we have defined $\lambda \left( \Gamma \right) \overset{%
\text{def}}{=}\left( 4\Gamma \right) /h$. Of the four cases we have
considered, this final case exhibits the coolest optimum paths with the
slowest speed (lowest entropy production rate and lowest entropic speed).

With all four cases analyzed, we plot in Fig. $2$ the expressions we derived
for the geodesic paths from all four cases, starting from the initial
conditions $\theta _{0}=0$ and $\dot{\theta}_{0}=1$. Furthermore, we set $%
\Gamma /\hslash =1$ and $\lambda =2/\pi $. By\ imposing these assumptions in
Eqs. (\ref{geo1a}), (\ref{geo2}), (\ref{geo3}), and (\ref{geo4}), we obtain, 
\begin{equation}
\theta \left( \xi \right) =\xi \text{, }\theta \left( \xi \right) =\frac{\pi 
}{2}\sin ^{-1}\left( \frac{2}{\pi }\xi \right) \text{, }\theta \left( \xi
\right) =\frac{\pi }{2}\frac{\xi }{\frac{\pi }{2}-\xi }\text{, and }\theta
\left( \xi \right) =\frac{\pi }{2}\log \left( \frac{1}{1-\frac{2}{\pi }\xi }%
\right) \text{, }
\end{equation}%
for the constant case, the oscillatory case, the power-law case, and the
exponential case, respectively. We also plot, in Fig. $3$, the Fisher
information $\mathcal{F}(\theta )$ along with the constant values of the
entropic speed and entropy production rate for each of the four cases. Each
new case we considered exhibited cooler optimum paths, albeit with slower
entropic speed. This result is summarized by the following observation, 
\begin{equation}
0\leq e^{-\lambda \theta _{0}}\leq \frac{1}{\left( 1+\lambda \theta
_{0}\right) ^{2}}\leq \left\vert \cos \left( \lambda \theta _{0}\right)
\right\vert \leq 1\text{,}
\end{equation}%
when $\theta _{0}\in 
\mathbb{R}
_{+}$ and $\lambda \left( \Gamma \right) \overset{\text{def}}{=}\left(
4\Gamma \right) /h\gg 1$. However, this ranking is not true for all values
of $\lambda $. In particular, when $0\leq \lambda \lesssim 1$, we find that
the entropic speed in the exponential scenario can outperform that of the
power-law scenario. To quantify this statement, note that Eqs. (\ref{speed3}%
) and (\ref{speed4}) imply, 
\begin{equation}
v_{\mathrm{E}}^{\left( \text{power-law}\right) }\left( \Gamma \right) =\frac{%
e^{\lambda \theta _{0}}}{\left( 1+\lambda \theta _{0}\right) ^{2}}v_{\mathrm{%
E}}^{\left( \text{power-law}\right) }\left( \Gamma \right) \text{.}
\end{equation}%
If we define the function $f_{\mathcal{P}}\left( \lambda \text{, }\theta
_{0}\right) $ as,%
\begin{equation}
f_{\mathcal{P}}\left( \lambda \text{, }\theta _{0}\right) \overset{\text{def}%
}{=}\frac{e^{\lambda \theta _{0}}}{\left( 1+\lambda \theta _{0}\right) ^{2}}%
\text{,}
\end{equation}%
then the values of $\lambda $ and $\theta _{0}$ for which this function is
negative correspond to values for which the exponential decay scenario gives
entropic speeds that are higher than the power-law decay scenario. We
formally express this as a parametric region $\mathcal{P}$ defined as, 
\begin{equation}
\mathcal{P}\overset{\text{def}}{=}\left\{ \left( \lambda \text{, }\theta
_{0}\right) \in 
\mathbb{R}
_{+}\times 
\mathbb{R}
_{+}:f_{\mathcal{P}}\left( \lambda \text{, }\theta _{0}\right) <0\right\} 
\text{,}  \label{region}
\end{equation}%
and plot this region in Fig. $4$. It is seen clearly that sufficiently large
values for $\lambda $ cause the power-law strategy to perform better than
the exponential strategy in terms of entropic speed. We note that a typical
value of $\lambda $ in experiments may be estimated by assuming a magnetic
field intensity $B_{\perp }$ of approximately $0.2$ \textrm{T} (typical of
neodymium magnets), and recalling that $\lambda =(4\Gamma )/h$ with $\Gamma
=(|e|\hslash B_{\perp })/(2mc)$. This yields a value $\lambda \simeq 37$ $[%
\mathrm{MKSA}]$.

Our findings are summarized in Table III, where we describe the behavior of
the Fisher information $\mathcal{F}(\theta )$, the entropic speed $v_{%
\mathrm{E}}$, and the entropy production rate $r_{\mathrm{E}}$ for each of
the four scenarios alongside a description of the Hamiltonian models that
produce them. We also list for each case the type of continuous-time quantum
search algorithm that resembles the quantum evolution.

\begin{table}[t]
\centering
\begin{tabular}{c|c|c|c|c}
\hline\hline
Quantum Search & Hamiltonian Model & Fisher Information & Speed & Entropy
Production Rate \\ \hline
Grover-like & original, constant $B_{\bot}$ & constant & higher & higher \\ 
Grover-like & generalized, oscillating $B_{\bot}$ & oscillatory & high & high
\\ 
fixed-point-like & generalized, power law decay of $B_{\bot}$ & power law
decay & low & low \\ 
fixed-point-like & generalized, exponential decay of $B_{\bot}$ & 
exponential decay & lower & lower \\ \hline
\end{tabular}%
\caption{Illustrative representation of the entropy production rate, speed,
and Fisher information in the chosen four $\mathrm{su}(2$; $\mathbb{C} )$
Hamiltonian models. Furthermore, for each model, we emphasize the
Grover-like or fixed-point-like property possessed by its corresponding
continuous time quantum search algorithm.}
\end{table}

\section{Concluding Remarks}

In what follows, we present a summary of main results, as well as
limitations and further investigations.

\subsection{Summary}

In this paper, we presented an information geometric characterization of
entropic speeds (see Eq. (\ref{vthermo})) and entropy production rates (see
Eq. (\ref{EPR})) that emerge from the geodesic motion (see Eq. (\ref{general}%
)) on manifolds of parametrized quantum states (see Eq. (\ref{output})).
These pure states emerge as outputs of suitable $\mathrm{su}\left( 2\text{; }%
\mathbb{C}
\right) $ time-dependent Hamiltonian evolutions (see Eq. (\ref{peter3}))
employed to specify distinct types of continuous-time quantum search
schemes. The Riemannian metrization on the manifold is specified by the
Fisher information (see Eq. (\ref{JIJ2})) evaluated along the parametrized
squared probability amplitudes (see Eqs. (\ref{1a}), (\ref{5b}), (\ref{4b}),
and (\ref{3a})) obtained from analysis of the quantum mechanical temporal
evolution of a spin-$1/2$ particle in an external time-dependent magnetic
field that prescribes the $\mathrm{su}\left( 2\text{; }%
\mathbb{C}
\right) $ Hamiltonian model. In Fig. $1$, we show the manner in which a
specific Fisher information behavior arises from a given $\mathrm{su}\left( 2%
\text{; }%
\mathbb{C}
\right) $ Hamiltonian model characterized by a particular magnetic field
configuration (see Table II). Using a minimum action principle to transfer a
quantum system from an initial state to a final state on the manifold in a
finite temporal interval\textbf{, }we demonstrate that the minimizing
(optimum) path is the shortest (geodesic, see Fig. $2$) path between the two
states and in particular, also minimizes the total entropy production, that
is, the thermodynamic divergence of the path in Eq. (\ref{divergence2}) that
occurs during the transfer (see Eqs. (\ref{general}) and (\ref{general1})).
Then, by evaluating the entropic speed and the total entropy production
along the optimum transfer paths (see Eqs. (\ref{geo1a}), (\ref{geo2}), (\ref%
{geo3}), and (\ref{geo4})) in the four chosen physical scenarios of interest
in analog quantum search problems, we demonstrate in a clear quantitative
manner that to a faster transfer there necessarily corresponds a higher
entropy production rate (see Fig. $3$ and Table III). We therefore conclude
that lower (entropic) efficiency values do appear to accompany higher
(entropic) speed values in quantum transfer processes.

In particular, the probability paths with an exponentially decaying Fisher
information seem to achieve the highest entropic efficiency (see Eq. (\ref%
{lambda4})) with the cost of also having the lowest entropic speed (see Eq. (%
\ref{speed4})). By contrast, the probability paths with a constant Fisher
information appear to be the fastest (see Eq. (\ref{speed1})) but also the
most inefficient from an entropic standpoint (see Eq. (\ref{lambda1})). A
graphical summary of these results, including all four $\mathrm{su}\left( 2%
\text{; }%
\mathbb{C}
\right) $ Hamiltonian models considered in this paper (see Table II), appear
in Fig. $3$ , Fig. $4$, and Table III.

\subsection{Limitations and Further Investigations}

In what follows, we report some limitations and insights that emerge from
our information geometric analysis.

\begin{enumerate}
\item[1)] First, we focused our information geometric analysis on quantum
mechanical evolutions between perfectly distinguishable initial and final
quantum states. Specifically, we have assumed a vanishing quantum overlap
in\ Eq. (\ref{good})\textbf{, }meaning that the initial source state is
exactly orthogonal to the final target state. Furthermore, we have also
assumed throughout our investigation the validity of the generalized
on-resonance condition as defined in\ Eq. (\ref{GRC}). Clearly, we expect
that scale effects and more important computational complications of
transition probabilities can emerge when departing from the on-resonance
condition. From a physical perspective, we also expect to find a richer
dynamics where the role of essential control field is not limited to $\vec{B}%
_{\perp }$\textbf{. }For this reason\textbf{\ }we expect that $\vec{B}%
_{\parallel }$ as well as the phase $\phi _{\omega }$ will play a key role
in our information geometric analysis. As pointed out by Rabi and
collaborators in\ Ref. \cite{rabi54}, the magnitude of $\vec{B}_{\perp }$ is
generally much smaller than the magnitude of $\vec{B}_{\parallel }$ in a
variety of nuclear resonance experiments. Interestingly, when comparing the
Farhi-Gutmann analog quantum search Hamiltonian \cite{farhi98} with Rabi's
original two-level quantum system Hamiltonian \cite{sakurai94}, the quantum
overlap $x$ between the initial and final states can be recast in terms of
the magnetic field intensities $B_{\perp }$ and $B_{\parallel }$ \cite%
{alsing19},%
\begin{equation}
x\left( B_{\perp }\text{, }B_{\parallel }\right) \overset{\text{def}}{=}%
\frac{B_{\parallel }}{B_{\perp }}\frac{1}{\sqrt{1+\left( \frac{B_{\parallel }%
}{B_{\perp }}\right) ^{2}}}\text{.}  \label{overlap}
\end{equation}%
Therefore, scenarios with $B_{\perp }\gg B_{\parallel }$ would correspond to
cases in which $0<x\ll 1$. Curiously, in the framework of quantum mechanical
spin-$1/2$ particle manipulations with magnetic fields, the so-called
strong-driving regime occurs when the Rabi frequency $\Omega _{\text{Rabi}}%
\overset{\text{def}}{=}\Gamma /\hslash \propto B_{\perp }$ is greater than
the Larmor frequency $\Omega _{\text{Larmor}}\overset{\text{def}}{=}\Delta
E/\hslash \propto B_{\parallel }$ with $\Delta E$ being the energy splitting
between the system's states \cite{london14,shim14}. In nuclear magnetic
resonance experiments performed in the strong-driving regime, the intensity
of the static magnetic field $\vec{B}_{\parallel }$ can be less than a few
micro-tesla ($B_{\parallel }\simeq 10^{-6}$ $\mathrm{T}$, with $1$ $\mathrm{T%
}=10^{4}$ \textrm{G} ) \cite{shim14}. In the ultrastrong-driving regime
where $B_{\perp }\gg B_{\parallel }$ for instance, the \textquotedblleft
on-resonance\textquotedblright\ driving does not appear to be a good control
strategy in terms of total evolution time and/or energy-type cost functional
minimization \cite{hirose18}. Despite the aforementioned limitations of our
analysis, we truly think the information geometric approach proposed in this
paper deserves further investigation since it could be leveraged to address
a very important problem in quantum information processing, namely provide a
theoretical basis for a first of its kind quantitative classification scheme
for the various driving regimes in spin manipulations based upon both
thermodynamic efficiency and minimum evolution time. In view of these
arguments, it is our intention to extend our understanding of these topics
by considering both departures from perfect distinguishability and
generalized out-of-resonance conditions (for further details, see also Ref. 
\cite{carloquantum20}) in our future information geometric investigations.

\item[2)] Second, we emphasize that in the \ framework of finite-time
thermodynamics, the rate of entropy production $d\sigma /dt$ (that is, $r_{%
\mathrm{E}}$ in our terminology) is constant along an optimum path (that is,
optimal process trajectory) only for linear processes \cite{ries95}. These
processes are characterized by Onsager coefficients $\mathrm{L}_{ij}$ \cite%
{onsager31a,onsager31b} that do not depend on the fluxes $\dot{X}_{i}$, with
fluxes being the rate of change in time of the particular extensive
parameters $X_{i}$ of the thermodynamic system (volume and internal energy,
for instance). Departing from the linearity condition, it happens that \cite%
{ries95},%
\begin{equation}
\frac{d\sigma }{dt}+\dot{X}_{i}\frac{\partial \mathrm{R}_{ij}}{\partial \dot{%
X}_{k}}\dot{X}_{k}\dot{X}_{j}=\text{constant,}  \label{resistenza}
\end{equation}%
with $\mathrm{R}_{ij}$ being the coefficients of the so-called resistance
matrix, the inverse of the matrix of Onsager coefficients $\mathrm{L}_{ij}$.
From Eq. (\ref{general}), we note that if $\mathrm{R}_{ij}$ does not depend
on $\dot{X}\overset{\text{def}}{=}dX/dt$, then\textbf{\ }the rate of entropy
production $d\sigma /dt$ is constant. From an information geometric
viewpoint, the relaxation of the working assumption of thermodynamic
linearity would lead to a new geometrical setting where the information
tensor $g_{\alpha \beta }$ would not only depend on $\theta $ but also on $%
\dot{\theta}\overset{\text{def}}{=}d\theta /d\xi $. This fact in turn, would
result in a departure from the Riemannian geometric platform to what is
known as a Finslerian geometric setting \cite{maria98,clayton15}. We have
not addressed the consequences of such a possible transition in our paper
and it would be intriguing to investigate what would occur due to relaxation
of the linearity working assumption. Interestingly, within the framework of
minimum dissipation protocols in spin systems \cite{crooks15,crooks17}, the
components of Kirkwoods friction tensor \cite{kirkwood46} play the role of
the resistance metric coefficients that appear in\ Eq. (\ref{resistenza}).
We have limited our investigation in this paper to a single-parameter
analysis with a metric tensor that does not depend on the derivative of the
parameter itself. We hope to extend our analysis to multi-parameter and/or
flux-dependencies in future efforts by addressing further computational
challenges and gaining even deeper insights toward a realistic comparative
thermodynamical analysis of quantum search algorithms from an information
geometric perspective. Finally, for a discussion on the Finslerian
geometrization of Hamiltonian dynamics for physical systems with potentials
depending explicitly on time and velocities, we refer to Refs. \cite%
{dibari97,pettini07}.

\item[3)] Third, our work can be viewed as leading to several intriguing
foundational insights. From a physics perspective, given that analog quantum
search algorithms can be understood in terms of two-level quantum systems 
\cite{alsing19,byrnes18}, quantum searching can be linked to the study of
the quantum mechanical evolution of an electron in an external magnetic
field. From a classical electrodynamic perspective, an electric charge $e$
moving in a magnetic field $\vec{B}$ is deflected by a force $\vec{f}$
proportional to the product of the charge $e$ and velocity $\vec{v}$,%
\begin{equation}
\vec{f}=e\frac{\vec{v}}{c}\times \vec{B}\text{.}  \label{force}
\end{equation}

Therefore, since the force depends upon the velocity, an electron in an
external magnetic field is an example of a conservative dynamical system
which is not macroscopically reversible \cite{onsager31a,onsager31b}.
Examples of irreversible transport processes are heat conduction, electrical
conduction, and \emph{diffusion}. Interestingly, Grover's search algorithm
was initially invented and presented in terms of a Markov \emph{diffusion}
process \cite{grover01}. Therefore, a potentially fundamental physical
insight arising from our information geometric investigation of minimum
entropy product paths from quantum mechanical evolutions corresponding to
two-level quantum systems mimicking analog quantum search Hamiltonian
evolutions is that a realistic thermodynamic analysis of quantum search
algorithms will eventually emerge from exploring the mathematical analysis
of Markov processes \cite{schnakengerg76,ge10}. At this stage, such a
statement is purely speculative. In order to provide a serious theoretical
analysis of both fast and thermodynamically efficient quantum search
schemes, we require a deeper understanding of the role played by
fluctuations and dissipation \cite{rondoni08,esposito10} in irreversible
transport processes together with their connection to the physical mechanism
underlying analog quantum search algorithms within realistic models of
computation.
\end{enumerate}

In conclusion, despite its current limitations, we view our investigation
presented in this paper as a natural progression of our work presented in
Refs. \cite{cafaropre18,carlopre20}. It constitutes a nontrivial preliminary
effort toward understanding quantum search algorithms from a thermodynamical
perspective developed within an information geometric setting. It is our
intention to improve upon the analysis provided in this paper and pursue
these fascinating avenues of investigation in forthcoming scientific
efforts. Of course, it is our sincere hope that our work will inspire other
scientists to further explore these research avenues in the near future.

\begin{acknowledgments}
C. C. is grateful to the United States Air Force Research Laboratory (AFRL)
Summer Faculty Fellowship Program for providing support for this work. Any
opinions, findings and conclusions or recommendations expressed in this
manuscript are those of the authors and do not necessarily reflect the views
of AFRL.
\end{acknowledgments}


\begin{thebibliography}{999}
\bibitem{frey} M. R. Frey,\emph{\ Quantum speed-limits-primer, perspectives,
and potential future directions}, Quantum Information Processing \textbf{15}%
, 3919 (2016).

\bibitem{nature} D. Castelvecchi, \emph{Clash of the physical laws}, Nature 
\textbf{543}, 597 (2017).

\bibitem{renner18} P. Faist and R. Renner, \emph{Fundamental work cost of
quantum processes}, Phys. Rev. \textbf{X8}, 021011 (2018).

\bibitem{deffner17} S. Campbell and S. Deffner, \emph{Trade-off between
speed and cost in shortcuts to adiabaticity}, Phys. Rev. Lett. \textbf{118},
100601 (2017).

\bibitem{wootters} W. K. Wootters, \emph{Statistical distance and Hilbert
space}, Phys. Rev. \textbf{D23}, 357 (1981).

\bibitem{landau} L. D. Landau and E. M. Lifshitz, \emph{Statistical Physics}%
, Pergamon (1977).

\bibitem{ruppeiner79} G. Ruppeiner, \emph{Thermodynamics: A Riemannian
geometric model}, Phys. Rev. \textbf{A20}, 1608 (1979).

\bibitem{ruppeiner95} G. Ruppeiner, \emph{Riemannian geometry in
thermodynamic fluctuation theory}, Rev. Mod. Phys. \textbf{67}, 605 (1995).

\bibitem{salamon85} P. Salamon, J. D. Nulton, and R. S. Berry, \emph{Length
in statistical thermodynamics}, J. Chem. Phys. \textbf{82}, 2433 (1985).

\bibitem{crooks07} G. E. Crooks,\emph{\ Measuring thermodynamic length},
Phys. Rev. Lett. \textbf{99}, 100602 (2007).

\bibitem{caves94} S. L. Braunstein and C. M. Caves, \emph{Statistical
distance and geometry of quantum states}, Phys. Rev. Lett. \textbf{72}, 3439
(1994).

\bibitem{cover06} T. M. Cover and J. A. Thomas, \emph{Elements of
Information Theory}, John Wiley \& Sons, Inc. (2006).

\bibitem{amari00} S.\ Amari and H. Nagaoka, \emph{Methods of Information
Geometry}, Oxford University Press (2000).

\bibitem{felice18} D. Felice, C. Cafaro, and S. Mancini, \emph{Information
geometric methods for complexity}, Chaos \textbf{28}, 032101 (2018).\emph{\ }

\bibitem{salamon83} P. Salamon and R. S. Berry, \emph{Thermodynamic length
and dissipated availability}, Phys. Rev.\ Lett. \textbf{51}, 1127 (1983).

\bibitem{salamon89} K. H. Hoffmann, B. Andresen, and P. Salamon, \emph{%
Measures of dissipation}, Phys. Rev.\ \textbf{A39}, 3618 (1989).

\bibitem{ries95} W. Spirtkl and H. Ries, \emph{Optimal finite-time
endoreversible processes}, Phys. Rev. \textbf{E52}, 3485 (1995).

\bibitem{diosi96} L. Diosi, K. Kulacsy, B. Lukacs, and A. Racz, \emph{%
Thermodynamic length, time, speed, and optimum path to minimize entropy
production}, J. Chem. Phys. \textbf{105}, 11220 (1996).

\bibitem{diosi00} L. Diosi and P. Salamon, \emph{From statistical distances
to minimally dissipative processes}, in Thermodynamics of Energy Conversion
and Transport, edited by S. Sieniutycz and A. De Vos (Springer, New York,
2000), pp. 286-318.

\bibitem{sivak12} D. A. Sivak and G. E. Crooks, \emph{Thermodynamic metrics
and optimal paths}, Phys. Rev. Lett. \textbf{108}, 190602 (2012).

\bibitem{crooks12b} D. A. Sivak and G. E. Crooks, \emph{Near-equilibrium
measurements of nonequilibrium free energy}, Phys. Rev. Lett. \textbf{108},
150601 (2012).

\bibitem{crooks15} G. M. Rostskoff and G. E. Crooks, \emph{Optimal control
in nonequilibrium systems: Dynamic Riemannian geometry of the Ising model},
Phys. Rev. \textbf{E92}, 060102(R) (2015).

\bibitem{crooks17} G. M. Rostskoff, G. E. Crooks, and E. Vanden-Eijnden, 
\emph{Geometric approach to optimal nonequilibrium control: Minimizing
dissipation in nanomagnetic spin systems}, Phys. Rev. \textbf{E95}, 012148
(2017).

\bibitem{jaynes57a} E. T. Jaynes, \emph{Information theory and statistical
mechanics. I}, Phys. Rev. \textbf{106}, 620 (1957).

\bibitem{jaynes57b} E. T. Jaynes, \emph{Information theory and statistical
mechanics. II}, Phys. Rev. \textbf{108}, 171 (1957).

\bibitem{frieden98} B. R. Frieden, \emph{Physics from Fisher Information},
Cambridge University Press (1998).

\bibitem{frieden99} B. R. Frieden, A. Plastino, A. R. Plastino, and B. H.
Soffer, \emph{Fisher-based thermodynamics: Its Legendre transform and
concavity properties}, Phys. Rev. \textbf{E60}, 48 (1999).

\bibitem{frieden02} B. R. Frieden,\ A. Plastino, A. R. Plastino, and B. H.
Soffer, \emph{Schr\"{o}dinger link between nonequilibrium thermodynamics and
Fisher information}, Phys. Rev. \textbf{E66}, 046128 (2002).

\bibitem{grover97} L. K. Grover, \emph{Quantum\ mechanics helps in searching
for a needle in a haystack}, Phys. Rev. Lett. \textbf{79}, 325 (1997).

\bibitem{nielsenbook} M.\ A. Nielsen and I. L. Chuang, \emph{Quantum
Computation and Information}, Cambridge University Press (2000).

\bibitem{alvarez00} J. J. Alvarez and C. Gomez, \emph{A comment on Fisher
information and quantum algorithms}, arXiv:quant-ph/9910115 (2000).

\bibitem{wadati01} A. Miyake and M. Wadati, \emph{Geometric strategy for the
optimal quantum search}, Phys. Rev. \textbf{A64}, 042317 (2001).

\bibitem{cafaro2012A} C. Cafaro and S. Mancini, \emph{An information
geometric viewpoint of algorithms in quantum computing}, AIP Conf. Proc. 
\textbf{1443}, 374 (2012).

\bibitem{cafaro2012B} C. Cafaro and S. Mancini, \emph{On Grover's search
algorithm from a quantum information geometry viewpoint}, Physica \textbf{%
A391}, 1610 (2012).

\bibitem{cafaro2017} C. Cafaro, \emph{Geometric algebra and information
geometry for quantum computational software}, Physica \textbf{A470}, 154
(2017).

\bibitem{bennett82} C. H. Bennett, \emph{The thermodynamics of computation-\
A review}, Int. J. Theor. Phys. \textbf{21}, 905 (1982).

\bibitem{parrondo15} J. M. R. Parrondo, J. M. Horowitz, and T. Sagawa, \emph{%
Thermodynamics of information}, Nat. Phys. \textbf{11}, 131 (2015).

\bibitem{carlo13} C. Cafaro and P. van Loock, \emph{Towards an entropic
analysis of quantum error correction with imperfections}, in Bayesian
Inference and Maximum Entropy Methods in Science and Engineering, AIP Conf.
Proc. \textbf{1553}, 275 (2013).

\bibitem{carlo14b} C. Cafaro and P. van Loock, \emph{An entropic analysis of
approximate quantum error correction}, Physica \textbf{A404}, 34 (2014).

\bibitem{beals13} R. Beals,\ S. Brierley, O. Gray, A. W. Harrow,\ S. Kutin,
N. Linden, D. Shepherd, and M. Stather, \emph{Efficient distributed quantum
computing}, Proc. R. Soc. \textbf{A469}, 20120686 (2013).

\bibitem{perlner17} R. Perlner and Y.-K. Liu, \emph{Thermodynamic analysis
of classical and quantum search algorithms}, arXiv:quant-ph/1709.10510
(2017).

\bibitem{cafaropre18} C. Cafaro and P. M.\ Alsing, \emph{Decrease of Fisher
information and the information geometry of evolution equations for quantum
mechanical probability amplitudes}, Phys. Rev. \textbf{E97}, 042110 (2018).

\bibitem{byrnes18} T. Byrnes, G. Forster, and L. Tessler, \emph{Generalized
Grover's algorithm for multiple phase inversion states}, Phys. Rev. Lett. 
\textbf{120}, 060501 (2018).

\bibitem{alsing19} C. Cafaro and P. M. Alsing, \emph{Continuous-time quantum
search and time-dependent two-level quantum systems}, Int. J. Quantum
Information \textbf{17}, 1950025 (2019).

\bibitem{alsing19b} C. Cafaro and P. M. Alsing, \emph{Theoretical analysis
of a nearly optimal analog quantum search}, Physica Scripta \textbf{94},
085103 (2019).

\bibitem{carlopre20} C. Cafaro and P. M.\ Alsing, \emph{Information geometry
aspects of minimum entropy production paths from quantum mechanical
evolutions}, Physical Review \textbf{E101}, 022110 (2020).

\bibitem{carloquantum20} C. Cafaro, S. Gassner, and P. M. Alsing, \emph{%
Information geometric perspective on off-resonance effects in driven
two-level quantum systems}, Quantum Reports \textbf{2}, 166 (2020).

\bibitem{steven20} S. Gassner, C. Cafaro, and S. Capozziello, \emph{%
Transition probabilities in generalized quantum search Hamiltonian evolutions%
}, Int. Journal of Geometric Methods in Modern Physics \textbf{17}, 2050006
(2020).

\bibitem{brau996} S. L. Braunstein, C. M. Caves, and G. J. Milburn, \emph{%
Generalized uncertainty relations: theory, examples, and Lorenz invariance,}
Annals of Physics \textbf{247}, 135 (1996).

\bibitem{crooks12} G. E. Crooks, \emph{Fisher information and statistical
mechanics}, Technical Note 008v4, available at http://threeplusone.com/sher
(2012).

\bibitem{luo03} S. Luo, \emph{Wigner-Yanase skew information and uncertainty
relations}, Phys. Rev. Lett. \textbf{91}, 180403 (2003).

\bibitem{luo06} S.-L. Luo, \emph{Fisher information of wavefunctions:
Classical and quantum}, Chin. Phys. Lett. \textbf{23}, 3127 (2006).

\bibitem{boixo07} S. Boixo,\ S. T. Flammia, C. M. Caves, and JM Geremia, 
\emph{Generalized limits for single parameter quantum estimation}, Phys.
Rev. Lett. \textbf{98}, 090401 (2007).

\bibitem{taddei13} M. M. Taddei, B. M. Escher, L. Davidovich, and R. L. de
Matos Filho, \emph{Quantum speed limit for physical processes}, Phys. Rev.
Lett. \textbf{110}, 050402 (2013).

\bibitem{pires16} D. Paiva Pires, M. Cianciaruso, L. C. Celeri, G. Adesso,
and D. O. Soares-Pinto, \emph{Generalized geometric quantum speed limits},
Phys. Rev. \textbf{X6}, 021031 (2016).

\bibitem{bures69} D. J. C. Bures, \emph{An extension of Kakutani's theorem
on infinite product measures to the tensor product of semifinite w}$^{\ast}$%
\emph{-algebras}, Trans. Am. Math. Soc. \textbf{135}, 199 (1969).

\bibitem{uhlmann76} A. Uhlmann, \emph{The transition probability in the
state space of a }$\ast$\emph{-algebra}, Rep. Math. Phys. \textbf{9}, 273
(1976).

\bibitem{ali16} C. Cafaro and S. A. Ali, \emph{Maximum caliber inference and
the stochastic Ising model}\textbf{, }Phys. Rev. \textbf{E94}, 052145 (2016).

\bibitem{zanardi07} P. Zanardi, L. Campos Venuti, and P. Giorda, \emph{Bures
metric over thermal state manifolds and quantum criticality}, Phys. Rev. 
\textbf{A76}, 062318 (2007).

\bibitem{caticha12} A. Caticha, \emph{Entropic Inference and the Foundations
of Physics}; USP Press: S\~{a}o Paulo, Brazil (2012); Available online:
http://www.albany.edu/physics/ACaticha-EIFP-book.pdf.

\bibitem{weinhold75} F. Weinhold, \emph{Metric geometry of equilibrium
thermodynamics}, J. Chem. Phys. \textbf{63}, 2479 (1975).

\bibitem{salamon84} P. Salamon, J. Nulton, and E. Ihrig, \emph{On the
relation between entropy and energy versions of thermodynamic length}, J.
Chem. Phys. \textbf{80}, 436 (1984).

\bibitem{felice90} F. De Felice and J. S. Clarke, \emph{Relativity on curved
manifolds}, Cambridge University Press (1990).

\bibitem{salamon88} P. Salamon, J. D. Nulton, J. R. Harland, J. Pedersen, G.
Ruppeiner, and L. Liao, \emph{Simulated annealing with constant
thermodynamic speed}, Computer Physics Communications \textbf{49}, 423
(1988).

\bibitem{andresen94} B. Andresen and J. M. Gordon, \emph{Constant
thermodynamic speed for minimizing entropy production in thermodynamic
processes and simulated annealing}, Phys. Rev. \textbf{E50}, 4346 (1994).

\bibitem{pezze09} L. Pezz\'{e} and A. Smerzi, \emph{Entanglement, nonlinear
dynamics, and the Heisenberg limit}, Phys. Rev. Lett. \textbf{102}, 100401
(2009).

\bibitem{beretta05} E. P. Gyftopoulos and G. P. Beretta, \emph{%
Thermodynamics: Foundations and Applications}, Dover Publications, Inc.
(2005).

\bibitem{salamon80A} P. Salamon, A. Nitzan, B. Andresen, and R. S. Berry, 
\emph{Minimum entropy production and the optimization of heat engines},
Phys. Rev. \textbf{A21}, 2115 (1980).

\bibitem{tasaki16} N. Shiraishi, K. Saito, and H. Tasaki, \emph{Universal
trade-off relation between power and efficiency for heat engines}, Phys.
Rev. Lett. \textbf{117}, 190601 (2016).

\bibitem{seifert18} P. Pietzonka and U. Seifert, \emph{Universal trade-off
between power, efficiency, and constancy in steady-state heat engines},
Phys. Rev. Lett. \textbf{120}, 190602 (2018).

\bibitem{anandan90} J. Anandan and Y. Aharonov, \emph{Geometry of quantum
evolution}, Phys. Rev. Lett.\textbf{\ 65}, 1697 (1990).

\bibitem{sakurai94} J. J. Sakurai, \emph{Modern Quantum Mechanics},
Addison-Wesley Publishing Company, Inc. (1994).

\bibitem{carlopra10} C. Cafaro and S. Mancini, \emph{Quantum stabilizer
codes for correlated and asymmetric depolarizing errors}, Phys. Rev. \textbf{%
A82}, 012306 (2010).

\bibitem{carlopra14} C. Cafaro and P. van Loock, \emph{Approximate quantum
error correction for generalized amplitude-damping errors}, Phys. Rev. 
\textbf{A89}, 022316 (2014).

\bibitem{landau32} L. D. Landau, \emph{A theory of energy transfer. }$II$,
Phys. Z. Sowjet. \textbf{2}, 46 (1932).

\bibitem{zener32a} C. Zener, \emph{Non-adiabatic crossing of energy levels},
Proc. R.\ Soc. London,\ Ser. \textbf{A137}, 696 (1932).

\bibitem{rabi37} I. I. Rabi, \emph{Space quantization in a gyrating magnetic
field}, Phys. Rev. \textbf{51}, 652 (1937).

\bibitem{rabi54} I. I. Rabi, N. F. Ramsey, and J. Schwinger, \emph{Use of
rotating coordinates in magnetic resonance problems}, Rev. Mod. Phys. 
\textbf{26}, 167 (1954).

\bibitem{zener32} N. Rosen and C. Zener, \emph{Double Stern-Gerlach
experiment and related collision phenomena}, Phys. Rev. \textbf{40}, 502
(1932).

\bibitem{barnes12} E. Barnes and S. Das Sarma, \emph{Analytically solvable
driven time-dependent two-level quantum systems}, Phys. Rev. Lett. \textbf{%
109}, 060401 (2012).

\bibitem{messina14} A. Messina and H. Nakazato, \emph{Analytically solvable
Hamiltonians for quantum two-level systems and their dynamics}, J. Phys. A:
Math. Theor. \textbf{47}, 445302 (2014).

\bibitem{grimaudo18} R. Grimaudo, A. S. M. de Castro, H. Nakazato, and A.
Messina, \emph{Classes of exactly solvable generalized semi-classical Rabi
systems}, Ann. Phys. (Berlin) \textbf{2018}, 1800198.

\bibitem{bloch46} F. Bloch, \emph{Nuclear induction}, Phys. Rev. \textbf{70}%
, 460 (1946).

\bibitem{bloch46b} F. Bloch, W. W. Hansen, and M. Packard, \emph{The nuclear
induction experiment}, Phys. Rev. \textbf{70}, 474 (1946).

\bibitem{bloch53} R. K. Wangness and F. Bloch, \emph{The dynamical theory of
nuclear induction}, Phys. Rev.\textbf{\ 89}, 728 (1953).

\bibitem{grover05} L. K. Grover, \emph{Fixed-point quantum search}, Phys.
Rev. Lett. \textbf{95}, 150501 (2005).

\bibitem{farhi98} E. Farhi and S. Gutmann, \emph{An analog analogue of a
digital quantum computation}, Phys. Rev. \textbf{A57}, 2403 (1998).

\bibitem{dalzell17} A. M. Dalzell, T. J. Yoder, and I. L. Chuang,\emph{\
Fixed-point adiabatic quantum search}, Phys. Rev. \textbf{A95}, 012311
(2017).

\bibitem{bae02} J. Bae and Y. Kwon, \emph{Generalized quantum search
Hamiltonians}, Phys. Rev. \textbf{A66}, 012314 (2002).

\bibitem{roland02} J. Roland and N. J. Cerf, \emph{Quantum search by local
adiabatic evolution}, Phys. Rev. \textbf{A65}, 042308 (2002).

\bibitem{romanelli07} A. Perez and A. Romanelli, \emph{Nonadiabatic quantum
search algorithms}, Phys. Rev. \textbf{A76}, 052318 (2007).

\bibitem{reginatto} M. Reginatto and M. J. W. Hall, \emph{Quantum theory
from the geometry of evolving probabilities}, AIP Conf. Proc. \textbf{1443},
96 (2012).

\bibitem{fisher25} R.\ A. Fisher, \emph{Theory of statistical estimation},
Proc. Cambridge Philos.\ Soc. \textbf{22}, 700 (1925).

\bibitem{linnik59} Y. V. Linnik, \emph{An information-theoretic proof of the
central limit theorem with the Lindeberg condition}, Theory Probab. Appl. 
\textbf{4}, 288 (1959).

\bibitem{mckean66} H. P. McKean Jr., \emph{Speed of approach to equilibrium
for Kac's caricature of a Maxwellian gas}, Arch. Rational Mech. Anal. 
\textbf{21}, 343 (1966).

\bibitem{toscani92} G. Toscani, \emph{New a priori estimates for the
spatially homogeneous Boltzmann equation}, Continuum Mech. Thermodyn. 
\textbf{4}, 81 (1992).

\bibitem{villani98a} C. Villani,\emph{\ Fisher information estimates for
Boltzmann's collision operator}, J. Math. Pures Appl. \textbf{77}, 821
(1998).

\bibitem{villani98b} C. Villani, \emph{On the spatially homogeneous Landau
equation for Maxwellian molecules}, Math. Models Methods Appl. Sci. \textbf{8%
}, 957 (1998).

\bibitem{villani00} C. Villani, \emph{Decrease of the Fisher information for
the Landau equation with Maxwellian molecules}, Math. Models Methods Appl.\
Sci. \textbf{10}, 153 (2000).

\bibitem{wick54} G. C. Wick, \emph{Properties of Bethe-Salpeter wave
functions}, Phys. Rev. \textbf{96}, 1124 (1954).

\bibitem{london14} P. London, P. Balasubramanian, B. Naydenon, L. O.
McGuinness, and F. Jelezko, \emph{Strong driving of a single spin using
arbitrarily polarized fields}, Phys. Rev. \textbf{A90}, 012302 (2014).

\bibitem{shim14} J. H. Shim, S.-J. Lee, K.-K. Yu, S.-M. Hwang, and K. Kim, 
\emph{Strong pulsed excitations using circularly polarized fields for
ultra-low field NMR}, J. Mag. Res. \textbf{239}, 87 (2014).

\bibitem{hirose18} M. Hirose and P. Cappellaro, \emph{Time-optimal control
with finite bandwidth}, Quantum information Processing \textbf{17}, 88
(2018).

\bibitem{onsager31a} L. Onsager, \emph{Reciprocal relations in irreversible
processes.} I, Phys. Rev. \textbf{37}, 405 (1931).

\bibitem{onsager31b} L. Onsager, \emph{Reciprocal relations in irreversible
processes}. II, Phys. Rev. \textbf{38}, 2265 (1931).

\bibitem{maria98} M. Di Bari and P. Cipriani,\emph{\ Geometry and chaos on
Riemannian and Finsler manifolds}, Planetary and Space Science \textbf{46},
1543 (1998).

\bibitem{clayton15} J. D. Clayton, \emph{On Finsler geometry and
applications in mechanics: Review and new perspectives},\ Adv. Math. Phys.
Vol. 2015, Article ID 828475, 11 pages (2015).

\bibitem{kirkwood46} J. G. Kirkwood, \emph{The statistical mechanical theory
of transport processes}, J. Chem. Phys. \textbf{14}, 180 (1946).

\bibitem{dibari97} M. Di Bari, D. Boccaletti, P. Cipriani, and G. Pucacco, 
\emph{Dynamical behavior of Lagrangian systems on Finsler manifolds}, Phys.
Rev. \textbf{E55}, 6448 (1997).

\bibitem{pettini07} M. Pettini, \emph{Geometry and Topology in Hamiltonian
Dynamics and Statistical Mechanics}, Springer (2007).

\bibitem{grover01} L. K. Grover, \emph{From Schr\"{o}dinger's equation to
the quantum search algorithm}, Am. J. Phys. \textbf{69}, 769 (2001).

\bibitem{schnakengerg76} J. Schnakengerg, \emph{Network theory of
microscopic and macroscopic behavior of master equation systems}, Rev. Mod.
Phys. \textbf{48}, 571 (1976).

\bibitem{ge10} H. Ge and H. Qian, \emph{Physical origin of entropy
production, free energy dissipation, and their mathematical representations}%
, Phys. Rev. \textbf{E81}, 051133 (2010).

\bibitem{rondoni08} U. Marini Bettolo Marconi, A. Puglisi, L. Rondoni, and
A. Vulpiani, \emph{Fluctuation-dissipation: Response theory in statistical
physics}, Phys. Rep. \textbf{461}, 111 (2008).

\bibitem{esposito10} M.\ Esposito and C. Van den Broeck, \emph{Three
detailed fluctuation theorems}, Phys. Rev. Lett. \textbf{104}, 090601 (2010).
\end{thebibliography}
\end{document}